\documentstyle[sprocl,epsf]{article}
\newcommand{\ewxy}[2]{\setlength{\epsfxsize}{#2}\epsfbox[10 60 640 570]{#1}}

\newcommand{\bee}{\begin{equation}}
\newcommand{\ee}{\end{equation}}
\newcommand{\beea}{\begin{eqnarray}}
\newcommand{\eea}{\end{eqnarray}}
\newcommand{\rme}{{\rm e}}

\begin{document}
\title{ NONPERTURBATIVE QUANTUM FIELD THEORY ON THE LATTICE}
\author{
THOMAS DeGRAND}
\address{Department of Physics,
 University of Colorado, Boulder CO 80309-390} 
\maketitle\abstracts{
These lectures provide an introduction to lattice methods for
nonperturbative studies of quantum field theories, with an
emphasis on
Quantum Chromodynamics. 
Lecture 1 (Ch. 2): gauge field basics; lecture 2 (Ch. 3):
Abelian duality with a lattice regulator;
 (Ch. 4) simple lattice intuition;
lecture 3 (Ch. 5): standard methods (and results) for hadron spectroscopy;
lecture 4 (Ch. 6): bare actions and physics;
lecture 5 (Ch. 7): two case studies, the mass of the glueball, and  $\alpha_s(M_Z)$.
}
\section{Introduction}
The physics of the 20th century is the physics of local gauge invariance.
Lattice gauge theories present an extreme limit, being systems with exact
 internal symmetries but no  (or very restricted)
space time symmetries.
These lectures are intended to be an introduction to
the lattice approach to quantum field theories in general and
to QCD in particular \cite{STANDARD}.  
The use of a lattice regulator is a
standard technique for studying strong coupling when space-time
symmetries can be regarded as secondary.
Many techniques of lattice have found a home or at least a word in
the vocabulary of  string. Some of these techniques might be useful to
you.
Lattice methods have been used to study  random surfaces 
in the context of models for quantum gravity, but I will not discuss that
subject here.

Lattice gauge theory was invented by Wilson \cite{KEN} in 1974.
For its first five years or so it was  a subject studied via analytic
calculations. Then in 1979 Creutz, Jacobs, and Rebbi \cite{CJR}
performed the first numerical studies of a lattice gauge theory
using Monte Carlo simulation, and over the next few years the field
was transformed into one in which numerical methods combined with analytic
methods to study nonperturbative behavior.
By the late 1980's the field was dominated by
 very professional large scale simulations of QCD on supercomputers.
In the last few years some members of the lattice community have
been developing techniques for reducing the size of numerical simulations.
These techniques rely heavily on modern work on the renormalization
group and effective actions. Some of them actually work:
some lattice calculations can now be done on work stations.
Some of the techniques may have a wider range of applicability than
lattice QCD.

Most of the theoretical ideas that this audience might find interesting are
from the pre-computer days of lattice gauge theory.
I will begin by describing how gauge symmetry   is realized on the lattice
and then show how one studies confinement in the strong coupling limit.
Next, I will describe strong-weak coupling duality in the context
of lattice models. (This is the only place where I know how to
 explicitly perform a duality transformation on the field variables.)
I will next describe how lattice methods are used to obtain qualitative
insight into the behavior of various model systems, without using the computer.
Then the subject will change to QCD. I will describe how one
performs Monte Carlo simulations, and show the state of the art
in lattice calculations of hadron spectroscopy.
In one lecture I will talk about the connection between
 bare actions, effective actions, and scaling, and how this connection
is being used to construct  ``improved actions'' for QCD. I will
present examples to
 show the extent to which they actually do improve.
Two case studies: glueballs, and the calculation of $\alpha_s(M_Z)$
from the lattice, will conclude the review.

\section{Gauge Field Basics }
\subsection{Beginnings}

The lattice is a cutoff which regularizes the
ultraviolet divergences of quantum field theories. As with any regulator,
it must be removed after renormalization. Contact with experiment only
exists in the continuum limit, when the lattice spacing is taken to zero.

We are drawn to lattice methods by our desire to study nonperturbative
phenomena.  Older regularization schemes are tied closely to perturbative
expansions: one calculates a process to some order in a coupling constant;
divergences are removed order by order in perturbation theory.  The lattice,
however, is a nonperturbative cutoff. Before a calculation begins,
all wavelengths less than a lattice spacing are removed.
Generally one cannot carry out analytical studies of a field theory
for physically interesting parameter values.  However, lattice
techniques lend themselves naturally to implementation on digital
computers, and one can perform more-or-less realistic simulations of
quantum field theories, revealing their nonperturbative structure, on
present day computers.  I think it is fair to say that little of the
 quantitative results about QCD
which have been obtained in the last decade, could  have been gotten 
without the use of numerical methods.

On the lattice we sacrifice Lorentz invariance but preserve all internal
symmetries, including local gauge 
invariance.  This preservation is important for nonperturbative physics. 
 For example, gauge invariance is a property of the continuum theory 
which is nonperturbative, so maintaining it as we pass to the lattice 
means that all of its consequences (including current conservation and 
renormalizability) will be preserved.

It is very easy to write down an action for
 scalar fields regulated by a  lattice.  One just replaces the
space-time coordinate $x_\mu$ by a set of integers $n_\mu$ ($x_\mu=an_\mu$,
where $a$ is the lattice spacing). 
Field variables $\phi(x)$ are defined on sites $\phi(x_n) \equiv \phi_n$,
 The action, an integral over
the Lagrangian, is replaced by a sum over sites
\bee
\beta S = \int d^Dx {\cal L} \rightarrow a^4 \sum_n 
{\cal L}(\phi_n) .\label{2.1}
\ee
and the generating functional for Euclidean Green's functions is
replaced by an ordinary integral over the lattice fields
\bee
Z = \int (\prod_n d \phi_n ) e^{-\beta S}. \label{2.2} 
\ee
Gauge fields are a little more complicated. They carry a
space-time index $\mu$ in addition to an internal symmetry index $a$
($A_\mu^a(x))$ and are associated with a path in space $x_\mu(s)$: a
particle traversing a contour in space picks up a phase factor
\bee
\psi \rightarrow P(\exp \ ig \int_s dx_\mu A_\mu) \psi
\ee
\bee
 \equiv U(s)\psi(x). \label{2.3}
\ee
$P$ is a path-ordering factor analogous to the time-ordering
operator in ordinary quantum mechanics. Under a gauge transformation $g$,
$U(s)$ is rotated at each end:
\bee
U(s) \rightarrow g^{-1}(x_\mu(s))U(s)g(x_\mu(0)). \label{2.4} 
\ee
These considerations led Wilson \cite{KEN} to formulate gauge fields
on a space-time lattice, as follows:

The fundamental variables are elements of the gauge  group $G$ which live
on the links of a $D$-dimensional lattice, connecting $x$ and $x+ \mu$:
$U_\mu(x)$, with $U_\mu(x+\mu)^\dagger = U_\mu(x)$
 \bee
U_\mu(n)= \exp (igaT^aA^a_\mu(n))  \label{2.5}
\ee
for $SU(N)$.
($g$ is the coupling, $a$ the lattice spacing, $A_\mu$ the vector potential, 
and $T^a$ is a group generator).

Under a gauge transformation link variables transform as 
\bee
U_\mu (x) \rightarrow V(x) U_\mu (x) V(x+ \hat \mu)^\dagger  \label{2.10}
\ee
and site variables as 
\bee
\psi(x) \rightarrow V(x) \psi(x) \label{2.11}
\ee
so the only gauge invariant operators we can use as order parameters are 
 matter fields connected by  oriented ``strings" of U's (Fig. 1a)
\bee
\bar \psi(x_1) U _\mu(x)U_\mu(x+\hat \mu)\ldots  \psi (x_2)  \label{2.13}
\ee
or closed  oriented loops of U's (Fig. 1b)
\bee
{\rm Tr} \ldots U _\mu(x)U_\mu(x+\hat \mu)\ldots \rightarrow
{\rm Tr} \ldots U_\mu(x)V^\dagger (x+ \hat \mu)V(x+ \hat \mu)
U_\mu(x+\hat \mu)\ldots  .\label{2.12}
\ee

An action is specified by recalling that the classical Yang-Mills
action involves the curl of $A_\mu$, $F_{\mu\nu}$.
Thus a lattice action ought to involve a product of
$U_\mu$'s around some closed contour. In fact, there is enormous
arbitrariness at this point. We are trying to write down a bare action.
So far, the only requirement we  want to
impose is gauge invariance, and that will be automatically satisfied
for actions built of powers of traces of U's around closed loops,
with arbitrary coupling constants.
If we assume that the gauge fields are smooth, we can expand the link
variables in a power series in  $gaA_\mu's$. For almost any closed loop, the
leading term in the expansion will be proportional to $F_{\mu\nu}^2$.
We might want our action to have the same normalization as the continuum action.
This would provide one constraint among the lattice coupling constants.

 The simplest  contour has a perimeter of four links. In $SU(N)$
\bee
\beta S={{2N} \over {g^2}}\sum_n \sum_{\mu>\nu}{\rm  Re \ Tr \ }
\big( 1 - U_\mu(n)U_\nu(n+\hat\mu)
U^\dagger  _\mu(n+\hat\nu) U^\dagger  _\nu(n) \big).  \label{2.6}
\ee
This action is called the ``plaquette action'' or the
``Wilson action'' after its inventor.

Note Elitzur's theorem: \cite{ELI}
  only gauge invariant 
quantities have nonzero expectation values.

Let us see how this action reduces to the standard continuum action.
Specializing to the U(1) gauge group,
\bee
S= {1 \over {g^2}}\sum_n \sum_{\mu  > \nu} 
{\rm Re \ }(1 - \exp(iga[A_\mu(n) +A_\nu(n + \hat \mu)
-A_\mu(x+\hat\nu)-A_\nu(n)])).\label{2.7}
\ee
The naive continuum limit is taken by assuming that the lattice spacing 
$a$ is small, and Taylor expanding
 \bee
A_\mu(n+\hat\nu) = A_\mu(n) + a \partial_\nu A_\mu(n) + \ldots \label{2.8}
\ee
so the action becomes
\beea
S &= {1 \over {g^2}}\sum_n \sum_{\mu > \nu}
1-{\rm Re \ }( \exp(iga[a(\partial_\nu A_\mu -\partial_\mu A_\nu) + O(a^2)])) 
\\
&={1 \over {4g^2}}a^4\sum_n \sum_{\mu\nu} g^2F_{\mu\nu}^2 + \ldots  \\
&= {1 \over 4}\int d^4 x F_{\mu\nu}^2  \\ \nonumber \label{2.9}
\eea
transforming the sum on sites back to an integral.

\begin{figure}
\centerline{\ewxy{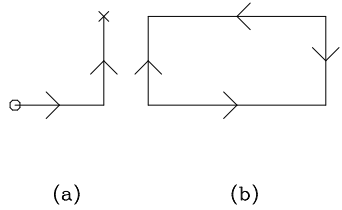}{80mm}
}
\caption{ Gauge invariant observables are
either  (a) ordered chains (``strings'')
of links connecting quarks and antiquarks or
(b) closed loops of link variables.}
\label{fig:figone}
\end{figure}

The best way people have found to
check for quark confinement is to ask the following question:
  In a world in which there are no light quarks, what is the 
potential $V(R)$ between a heavy $q\bar{q}$ pair?  If the limit as $R$ goes 
to infinity of $V(R)$ is infinite, we have confinement, if not, quarks are 
not confined. $V(R)$ can be computed by considering the partition function in Euclidean space for gauge fields in 
the presence of an external current distribution:
\bee
Z_J = \int [dU] \exp (-\beta S + i\int J_\mu A_\mu d^4 x) . \label{2.14} 
\ee
(for a non-Abelian gauge 
group insert color sums $J_\mu^a A_\mu^a$.)  If $J_\mu$ represents a 
point particle moving along a world line, it is a $\delta$-function on 
that world line (parameterized by $\ell_\mu$):
\bee
i \int J_\mu A_\mu d^4 x = i \oint A_\mu dl_\mu . \label{2.15} 
\ee
Gauge invariance, which implies current conservation, says that the 
current line cannot end, and so the loop is a closed loop.  Let's make 
it rectangular for simplicity, extending a distance R in some spatial 
direction and a distance T in the Euclidean time direction.  This is the 
famous Wilson loop. (See Fig. 2a.)

\begin{figure}
\centerline{\ewxy{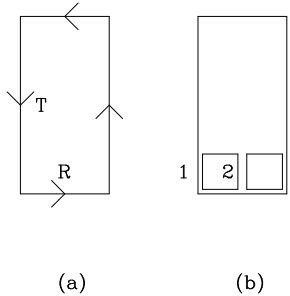}{80mm}
}
\caption{An $R \times T$ Wilson loop (a)
and the pattern of tiling which occurs when it is evaluated in strong
coupling (b). }
\label{fig:figthree}
\end{figure}

$Z_J$ describes the loop immersed in a sea of gluons. We can think of 
it as the partition function for a $q\bar{q}$ pair which is created at 
$t=0$, pulled apart a distance R, and then allowed to annihilate at a time 
T later.
 
To find the pair's energy, differentiate $\ln (Z)$ with respect to T 
(physically, we are measuring the response of the system as we stretch 
the loop a bit).  \cite{DOUGSHOW}
 But this is not quite right.  This calculation
 will include the 
vacuum energy of the gluons, which diverges.  What we really want is 
the change in energy of the system when we include the pair and when it 
is not present:
\beea
E_{q \bar q} &= -{\partial \over {\partial T}}
[ \ln Z_J - \ln Z_{J=0}] \\
&=   -{\partial \over {\partial T}} \ln ( Z_J / Z_{J=0})  \\
&= -{\partial \over {\partial T}} \ln{
{\int [dU]e^{-\beta S}e^{i\int A \cdot dl} }\over
{\int [dU]e^{-\beta S}} } \\
&= -{\partial \over {\partial T}} \ln\langle W \rangle_{J=0}  \label{2.16}
\eea
where $\langle W \rangle $
 is the expectation value of the Wilson loop in the background 
gluon field.  Thus the behavior of $\langle W \rangle$
 will tell us whether we have 
confinement or not.
 
For example, if 
$\langle W \rangle  \simeq \exp(-m(2R+2T)) $
(a so-called ``perimeter law") then $E=2m,$ and $m$ 
can be interpreted as  the quark mass.  
The energy of the pair is independent of R so that the quarks can 
separate arbitrarily far apart.  However, if 
$\langle W \rangle  \simeq \exp(-\sigma RT) $
(an ``area law"), $E=\sigma R$, and quarks are confined with a linear 
confinement potential.  The parameter $\sigma$ is called the ``string 
tension".
 In general, $\ln W$  \ is complicated:
$ -{\rm ln}\langle W \rangle = \sigma RT +2m(R+T) 
+ {\rm constant \ } + T/R + \ldots $
 Nevertheless, any area law term will dominate at large R and T and the 
theory will confine.
Notice that this confinement test will fail in the presence of light dynamical
fermions. As we separate the heavy $Q\bar Q$ pair, at some point it will become
energetically favorable to pop a light $\bar q q$ pair out of the vacuum,
so that we are separating two color singlet mesons. The Wilson loop
will show a perimeter law behavior.

 Now let us give an explicit demonstration of confinement for U(1) gauge 
theory. \cite{ITZY}  The Haar measure  for $U(1)$ is $(\theta_\mu = ga A_\mu)$
\bee
\int dU = \int_{-\pi}^\pi {{d \theta} \over {2\pi}} \label{2.17} 
\ee
and
\bee
Z= \int [dU] \exp (-\beta S)  \label{2.18}
\ee
where S is given by Eq. \ref{2.6}.
 
Strong coupling corresponds to small $\beta$.  Now 
we write down two trivial (but useful) results:
\beea
\int_{-\pi}^\pi {{d \theta} \over {2\pi}} e^{i(\theta + \phi)}
 &= 0 \\
\int_{-\pi}^\pi{{d \theta} \over {2\pi}} e^{i(\theta + \phi_1)}
e^{-i(\theta + \phi_2)} &=
e^{i(\phi_1-  \phi_2)}   \label{2.19} 
\eea
For small $\beta$ we may expand the exponential in 
Z as a power series in $\beta$:
\bee
Z= \int \prod_{links}{{d \theta} \over {2\pi}}\prod_{plaquettes}(1 + 
\beta({\rm Tr}U_p + {\rm h. \ c.}+ \ldots) 
\ee
\bee
= 1 + O(\beta^2) \label{2.20} 
\ee
since all linear terms vanish.
 
Let us compute the expectation value of an L $\times$ T Wilson loop 
(Fig. 2a):  There are LT plaquettes enclosed by it.
\bee
\langle W(L,T)\rangle =
{1 \over Z}\int \prod {{d \theta} \over {2\pi}} \exp (i \sum_{WL} \theta)
\prod_{plaquettes}(1 +\beta \cos  (\sum \theta) + \ldots)
\label{2.21} 
\ee
It's easy to convince one's self that $W$ is proportional to 
$\beta^{LT}$.  Pick one link on the boundary to integrate.  
We need a term 
in $\exp (-S)$ to contribute and cancel the phase of the boundary 
link appearing in the Wilson loop (link 1 of Fig. 2b).  Now we 
simultaneously need another 
O($\beta$) term to cancel the phases of link 2 (of Fig. 2b). 
 Only 
if we ``tile" the whole loop as shown do we get a nonzero expectation 
value; each plaquette contributes a factor $\beta$ so
\beea
 \langle W(L,T)\rangle  &\sim \beta^{LT} \\
E_{q \bar q} &= \sigma L \\
 \sigma &= - \ln \beta  \label{2.22}
\eea
This is confinement by linear potential.
 The calculation can be generalized to all gauge groups.
In the strong coupling limit, any lattice gauge theory shows confinement.
This calculation was first done by Wilson \cite{KEN} and was the first 
demonstration of quark confinement in a field theory in more than two
space-time dimensions, albeit in an unphysical limit.
This result had a lot to do with the early interest in lattice gauge theory:
one has an approximation in which confinement is automatic.

\subsection{Taking The Continuum Limit, and Producing a Number in MeV}

When we define a theory on a lattice the lattice spacing 
 $a$ is an ultraviolet cutoff and all the coupling constants
in the action are  the  bare couplings 
defined with respect it.
  When we take $a$ to zero we must also specify how 
$g(a)$ 
behaves.  The proper continuum limit comes when we take $a$ to zero 
holding physical quantities fixed, not when we take $a$ to zero holding 
the couplings fixed.
 
On the lattice, if all quark masses are set to zero,
 the only dimensionful parameter is the lattice spacing, 
so all masses scale like $1/a$. Said differently, one computes
the dimensionless combination $am(a)$. One can determine the lattice spacing by
fixing one mass from experiment. Then all other dimensionful quantities
can be predicted.

Now imagine computing some masses at several values of the lattice spacing.
(Pick several values of the bare parameters at random and
calculate masses for each set of couplings.)
Our calculated mass ratios will depend on the lattice cutoff.
The typical behavior will look like
\bee
(a m_1 (a))/(a m_2 (a)) = m_1(0)/m_2(0) + O(m_1a) + O((m_1 a)^2) +\dots \label{SCALING}
\ee
The leading term does not depend on the value of the UV cutoff, while the
 other terms  do.
The goal of a lattice calculation (like the goal of almost any calculation
in quantum field theory) is to discover the value of some physical observable
as the UV cutoff is taken to be very large, so the physics is in the first term.
Everything else is an artifact of the calculation.
We say that a calculation ``scales'' if the $a-$dependent terms in
Eq. \ref{SCALING} are zero or small enough that one can extrapolate
to $a=0$, and generically refer to all the $a-$dependent terms
as ``scale violations.''

We can imagine expressing  each dimensionless combination $am(a)$
as some function of the bare coupling(s) $\{g(a)\}$, $am = f(\{g(a))\}$.
 As $a\rightarrow 0$ we must tune the set of couplings $\{g(a)\}$ so
\bee
 \lim_{a \rightarrow 0}
{1 \over a} f(\{g(a)\}) \rightarrow {\rm constant} . \label{2.35}
\ee
From the point of view of the lattice theory,
 we must tune $\{g\}$ so that correlation lengths $1/ma$ diverge.  
This will occur only at the locations of second (or higher) order phase 
transitions in the lattice theory.   

Recall that the $\beta$-function is defined by
 \bee
 \beta(g)=a{{dg(a)} \over da} ={dg(a) \over d {\rm ln}(1/\Lambda a)}. 
\label{2.36}
\ee
(There is actually one equation for each coupling constant in the set.
$\Lambda$ is a dimensional parameter introduced to make the argument of 
the logarithm dimensionless.)  At a critical point $\beta(g_c)$ = 0.  
Thus the continuum limit is the limit
\bee
\lim_{a \rightarrow 0}\{g(a)\} \rightarrow \{g_c\} .\label{2.37}
\ee
Continuum QCD is a theory with one dimensionless coupling
constant.
In QCD  the fixed point is $g_c = 0 $
so we must tune the coupling to vanish as $a$ goes to zero.

Pushing this a little further, the
two-loop $\beta$-function is prescription independent,
\bee
\beta(g) = -b_1 g^3 +b_2 g^5 ,\label{2.38}
\ee
and so if we think that the lattice theory is reproducing the continuum,
and if we think that the coupling constant is small enough that
the two-loop beta-function is correct,
we might want to observe perturbative scaling, or
``asymptotic scaling", $m/\Lambda$ fixed, or $a$ varying with $g$ as
\bee
a \Lambda = ({1 \over {g^2(a)}})^{b_2 /( 2 b_1^2 )}\exp(- {1 \over {b_1 
 g^2(a)}})
 . \label{2.39}
\ee

Please note that asymptotic scaling is not scaling. Scaling means that
dimensionless ratios of physical observables do not depend on the
cutoff. Asymptotic scaling involves perturbation theory and the
definition of coupling constants. One can have one without the other.
(In fact, one can always define a coupling constant so that one
quantity shows asymptotic scaling.)

And this is not all.
There are actually two parts to  the problem of producing a number
 to compare with experiment.
One must first see scaling.
Then one needs to set the scale by
taking   some experimental number as input.
A complication that you may not have thought of
 is that the theory we simulate on the computer
is different from the real world. For example, 
a commonly used approximation is called the
``quenched approximation'': one neglects virtual quark loops, but includes
valence quarks in  the calculation.  The pion propagator is the propagator of
a $\bar q q$ pair, appropriately coupled, moving in a background of gluons.
This theory almost certainly does not have the same spectrum as QCD
with six flavors of dynamical quarks with their appropriate masses.
(In fact, an open question in the lattice community is, 
what is the accuracy of quenched approximation.)
Using one mass to set the scale from one of these
approximations to the real world might not give a prediction
for another mass which agrees with experiment.
We will see examples where this is important.

\section{ Abelian Duality with a Lattice Regulator}
``Every theorist knows'' that a phase with electrical confinement
(area law for Wilson loops, etc.) is has a dual description in terms of
magnetic variables which are screened, in analogy with the
Meissner effect in a superconductor. There are many sources for this
intuition. The one that people in my generation point to
are lattice calculations in Abelian gauge and spin models.
In these models one can explicitly perform the duality transformation.
In some models (notably the $d=2$ U(1) spin model and the $d=3$
gauge model) one can solve the dynamics sufficiently well to understand
the role of magnetic variables: the Kosterlitz-Thouless vortex-unbinding
transition in the former case, and Polyakov's demonstration of
confinement in the latter.
Classic references include  
  \cite{KOST}  \cite{JKKN}  \cite{BMK}  \cite{POLYAKOV}.

Let's begin with a $d=2$ $U(1)$ (or $O(2)$) spin model. On
each site of the lattice there is a unit-length spin $\vec s(x)$,
which can be labeled
by an angle (the orientation of the spin) $\phi(x)$. We assume nearest
neighbor spins interact  via ${\cal H} = 
\sum_x(1- \vec s(i) \cdot \vec s(i+\hat\mu)) =
\sum_{x,\mu}(1-\cos(\phi(x)-\phi(x+\hat\mu)))$. The partition function is
\bee
Z= \int_\pi^\pi [d\phi(x)] \exp(-\beta {\cal H}).
\ee
Let's perform the Fourier transform
\bee
\exp\big(-\beta(1-\cos(\theta))\big)= 
\ \sum_{-\infty}^\infty e^{in\theta}f_n(\beta)
\ee
with
\bee
f_n(\beta) \simeq {1 \over {\sqrt{2\pi \beta}}} e^{- n^2/(2\beta)}
\ee
to rewrite the partition function as
\bee
Z = \int [d\phi] \prod_{x,\mu} \sum_{n_\mu(x) = -\infty}^\infty
  e^{i\sum n_\mu(x)(\phi(x+\hat \mu)-\phi(x))} 
              e^{-\sum_{x,\mu}n_\mu(x)^2/(2\beta)}.
\ee
We can now integrate over the spin variables, which constrains
the $n_\mu's$:
\bee
\int d\phi(x) e^{i\phi(x)[n_\mu(x-\mu)-n_\mu(x)]}= \delta_{\nabla_\mu n_\mu,0}.
\ee
We can solve the constraint by introducing
\bee
n_\mu(x) = \epsilon_{\mu\nu}\nabla_\nu n(x)
\ee
which gives a partition function
\bee
Z= \prod_x  \sum_{n(x) = -\infty}^\infty \exp(
- \sum_{r,\mu}{{(\nabla_\mu n)^2} \over {2 \beta}}),
\ee
but at large $\beta$ the sums converge slowly. The Poisson resummation formula
\bee
\sum_l \delta(x-l) = \sum_m e^{2\pi i m x}
\ee
or
\bee
\sum_l g(l) = \sum_m \int d\phi g(\phi) e^{2\pi i m \phi}
\ee
allows us to rewrite the partition function as
\bee
Z= \prod_x \int [d\phi(x)] \sum_{m(x) = -\infty}^\infty \exp\big(
- \sum_{r,\mu}({{\nabla_\mu \phi)^2} \over {2 \beta}})
\exp(2 \pi i \sum_r m(r)\phi(r) \big)
\ee
or
\bee
Z = Z_{m=0}Z_{m\neq 0}
\ee
where
\bee
Z_{m=0} = \int [d\phi(x)] 
\exp(- \sum_{r,\mu}{{(\nabla_\mu \phi)^2} \over {2 \beta} })
\ee
is a free massless field theory (these are ``spin waves'' in
condensed-matter jargon), and the second term is a ``vortex component,''
\bee
Z_{m\neq 0} = \prod_x \sum_{m(x)=-\infty}^\infty 
\exp \big(-2\pi^2 \beta m(x) G(x-x')m(x') \big)
\ee
where
\bee
\nabla^2G = \delta(x-x').
\ee
That is, we have rewritten the original partition function in terms
of a set of variables which live on the sites, take on integer
values (0, $\pm 1$, $\pm 2, \dots$), and
interact by long-range Coulomb interactions. These variables are
called (depending on your dialect) vortices, defects, solitons, monopoles,
or instantons. Clearly, the $\beta \rightarrow \infty$ phase has
few vortices, and the low $\beta$ phase will have many.
This model has a phase transition at intermediate $\beta$ mediated
by the condensation of vortices  \cite{KOST}.

The algebra is similar in other dimensions. For $d=3$ the constraint is
\bee
\nabla_\mu n_\mu = 0 \rightarrow n_\mu = \epsilon_{\mu\nu\lambda}\nabla_\nu 
n_\lambda(x)
\ee
and if we impose a further ``gauge constraint'' 
\bee
\nabla_\lambda n_\lambda=0
\ee
we arrive at the partition function for a system of interacting, conserved,
oriented ``vortex strings''
\bee
Z(string)= \prod_{x,\mu} \sum_{m_\mu(x)=-\infty}^\infty
\exp\big(-2\pi^2\beta \sum_{x,x',\mu,\nu}
 m_\mu(x) G_{\mu,\nu}(x-x')m_\nu(x')\big)
|_{\nabla_\mu m_\mu=0}.
\ee
Again,
\bee
\nabla^2 G_{\mu\nu} = \delta_{\mu\nu}\delta(x)
\ee
and the oriented strings have a Coulomb  interaction.

U(1) gauge theories with the Wilson action
\bee
\cos(\theta_{\mu\nu}(x))=\cos(\theta_\mu(x)+\theta_\nu(x+\mu)-\theta_\mu(x+\nu)
-\theta_\nu(x))
\ee
can be treated similarly. The Fourier transform introduces a constraint
\bee
\exp(i l_{\mu\nu}\theta_{\mu\nu}) \rightarrow \delta_{\nabla_\mu l_{\mu\nu},0}.
\ee
In three dimensions the constraint is solved by setting
$l_{\mu\nu}=\epsilon_{\mu\nu\lambda}\partial_\lambda l$.
We have a system of interacting (charged) point defects: monopoles or
instantons.
In four dimensions the constraint is solved by setting
$l_{\mu\nu}=\epsilon_{\mu\nu\lambda}\partial_\lambda l_\sigma$
with $\partial_\sigma l_\sigma=0$. This is a system of interacting 
monopole world lines.
For more details about ``what comes next,'' see 
 \cite{BMK}  \cite{POLYAKOV}  \cite{TDDT}.

\subsection{Polyakov's Solution of 3-d U(1) Gauge Theory}
Polyakov showed that the $d=3$ U(1) gauge theory is permanently
confining for all values of the gauge coupling. 
The calculation begins by considering the grand partition
function for a gas of monopoles, where we restrict the possible 
charges to $\pm 1$.
(In keeping with the spirit of the
original demonstrations, all my numerical factors
are probably wrong.)
The monopole
fugacity $\zeta= \exp(-1/e^2)$. The partition function is
\bee
Z= \sum_N \sum_{q_J=\pm1} {\zeta^N \over N!} \int dx_1\dots dx_N
\exp(-\beta\pi^2 \sum_{a\ne b} {{q_a q_b}\over {|x_a-x_b|}})
\ee
or
\bee
Z= \int [d\chi]\exp(-{\pi \over \beta}\int dx (\nabla \chi)^2)
\sum_N\sum_{q=\pm1} 
{\zeta^N \over N!}dx_1\dots dx_N e^{i\sum_j q(x_j)\chi(x_j)}.
\ee
We can sum over the charges on a site 
$\sum_q \rightarrow \int dx(e^{i\chi}+e^{-i\chi})$
and exponentiate to produce a sine-Gordon Lagrangian
\bee
Z=\int [d\chi] \exp(-{\pi \over \beta}\int dx[ (\nabla\chi)^2 - M^2\cos \chi])
\ee
where $M^2 = \beta\zeta/\pi$.
Now let's introduce a Wilson loop into the calculation. We reverse
the derivation of the Wilson loop of the last lecture to write
\beea
W(C) = & \exp i \oint_C A \cdot dl \\
     = & \exp i \int d^4 x J_{ext} \cdot A \\
     = & \exp i \sum_r q(r) \eta(r) 
\eea
where $\eta$ is the  magnetic scalar potential arising from the
external current loop,
$\nabla^2 \eta = \vec \nabla \cdot \vec B_{ext} = \rho$.
We repeat the derivation including the external scalar potential
and find
\bee
Z(\eta) = \int [d\chi] \exp(-{\pi \over \beta}\int dx[ (\nabla(\chi-\eta))^2 - 
M^2\cos \chi]).
\ee
We solve the integral by steepest descent, and find
\bee
Z(\eta) = \exp\big(-{\pi \over \beta}\int dx[ (\nabla(\chi_{cl}-\eta))^2 - 
M^2\cos \chi_{cl}] \big)
\ee
where
\bee
\nabla^2 \chi_{cl} = M^2 \sin \chi_{cl} + \nabla^2 \eta
\ee
What is the solution? $\chi\neq 0$ only near the sheet and
 $\chi \simeq \exp(-M|z|)$ moving
transversely away from the sheet,
so $Z(\eta) \simeq \exp (-\gamma \times area)$ where
$\gamma \simeq \pi/\beta \times M^2 \times 1/M \simeq \exp(-1/e^2)$.
This is the signal of linear confinement. Hence the intuition:
electrical confinement is due to the free monopoles which screen
external magnetic fields.

\section{Simple Lattice Intuition}
\begin{figure}
\centerline{\ewxy{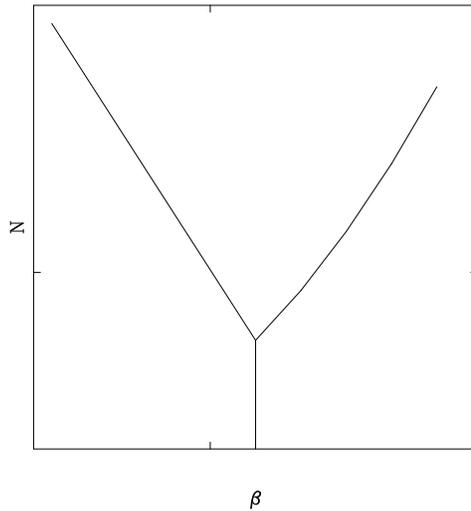}{80mm}
}
\caption{ Phase diagram for two-dimensional $Z(N)$ spin models.}
\label{fig:znfig}
\end{figure}

Let's finally consider some variations on these models.
$Z_N$ models are like $U(1)$ models, except the variables are restricted
to $\theta= 2\pi j/N$, $j=1\dots N$. They
can have three phases: a low-beta phase with many defects, a high-beta
magnetically-ordered or ``Higgs'' phase, and possibly a spin wave phase
at intermediate coupling.
The generic phase diagram is shown in Fig. \ref{fig:znfig}.
 It is easy to estimate
the location of the magnetic ordering transition: Fluctuations
in $\theta$ go like $ (\delta \theta)^2 \simeq
\langle \theta^2\rangle \simeq 1/\beta$, but
if this $\delta\theta < 1/N$ the spin cannot fluctuate from one allowed
value to another one, and the system freezes. This happens
at $\beta_c \simeq N^2$.

While the spin-wave phase is ``just free field theory,'' in two dimensions
the free behavior of the angular variable is
\bee
\langle \phi_i \phi_j\rangle \simeq {1\over \beta} \log({r_{ij} \over r_0}).
\ee
However, the spin-spin correlator is
\beea
\langle s_i \cdot s_j \rangle =  & \langle e^{i(\phi_i-\phi_j)}\rangle \\
                              = |r_{ij}|^{(1/\beta)} 
\eea
whose exponent varies continuously with the coupling. The continuous
variation in a critical index with an input parameter signals that
the system is always critical and one speaks of the
system as being characterized
by a ``line of critical points.''

We already saw that U(1) gauge models with a lattice
regulator confine in three Euclidean dimensions.  All lattice gauge theories
confine in two Euclidean dimensions. The demonstration is straightforward.
Write down the partition function and gauge fix so that all the links
in one direction (direction 2, in the following)
are set to the identity. (It may be necessary to
take open boundary conditions to do this.) The partition function
becomes
\bee
Z = \int \prod dU_1(x) \exp(\beta Tr U_1(x)U_1^\dagger(x+\hat 2)).
\ee
Now perform a second set of gauge transformations
$U_1(x) = U_1(x)U_1^\dagger(x+\hat 2)$, etc. The
partition function factorizes into a produce of one-plaquette actions
\bee
Z = [ \int dU \exp( \beta Tr U)]^{L \cdot T}.
\ee
The partition function has no singularities unless the single link
integral does (this can happen at large-N, for example). It is
easy to repeat the calculation for the Wilson loop and find an area law.

Monte Carlo simulations show that four-dimensional U(1) gauge theory
has a confining phase at strong coupling, apparently induced by monopole
condensation as in three dimensions, and a phase transition at $e^2 \simeq 1$
into a weak-coupling Coulomb phase. In that phase the
 relic monopole loops slightly
renormalize the electric charge. The order of the transition 
was controversial for many years, but  recent studies favor second order \cite{LANGU1}.
If it were first order, the theory would have
a continuum limit only in the Coulomb phase. If the transition is second
order, one could define  a continuum confined U(1) theory by approaching
the phase transition from within the confined phase.

Z(N) gauge theories in four dimensions have a phase diagram similar to
spin models. There is a strong-coupling confinement phase, a magnetically-ordered Higgs phase at large $\beta$, and at larger N,
an intermediate coupling Coulomb phase. When there are only two phases, the
deconfinement transition is first order.

There are many cases where the lattice can be used to provide insight into
a physical system, without the need for simulation. An example
of such a situation is the gauge-Higgs system, with a (schematic)
Lagrange density
\bee
{\cal L} = {\cal L}_{gauge} + (D_\mu\Phi)^2 - V(\Phi).
\ee
It is often useful to convert to radial and angular variables:
$\Phi = \rho \phi$ with $|\phi|^2=1$. Then
if we take $V(\Phi) =
\lambda(\Phi^2 - f^2)^2 = \lambda(\rho^2 - f^2)^2$, we can freeze out
the radial mode by taking the $\lambda
\rightarrow \infty$ limit, so $\langle \rho \rangle = f$.
Replacing the gradient by a lattice difference, one  can then write the
action in terms of two coupling constants as
\bee
{\cal L} = \beta_g Tr(1- UUUU) - \beta_h \phi_x^\dagger U_\mu(x) \phi(x+\mu).
\ee
The two coupling constants describe a plane of possible theories.
One studies the continuum behavior of these theories on the lattice
by finding all the critical points/lines in the phase diagram. One
recovers continuum physics by tuning couplings to approach those
critical points.

One can usually figure out the behavior of the theory on the boundaries
of the coupling constant plane without using Monte Carlo, and
use Monte Carlo only to explore the interior of the phase diagram.
This was a very common game in the days before the lattice community
became ``professional.''  Let's see how it works for a gauge-Higgs
model.

The first limiting case is $\beta_h=0$. This is a pure gauge theory,
which one has to analyze on its own.

Now $\beta_g=0$. This is a trivial limit, the strong-coupling confined
phase limit. One can see this in two ways. The first way is to realize that
there is no gauge interaction, so all the link
variables decouple and ${\cal L} = \phi^\dagger U \phi$ describes a set of 
random interactions. All $\langle \phi_i \phi_j\rangle$ correlators
vanish away from zero separation. Another way to see this limit is
to gauge fix to ``unitary gauge'' by rotating all the $\phi$'s to some
constant value $\phi_0$. Now
\bee
Z = \int \prod_{x,\mu} dU_\mu(x) \exp(\beta_h 
\sum_{x,\mu}\phi_0^\dagger U_\mu(x)
\phi_0)
\ee
and again the partition function factorizes in to a product of one-link
integrals.

At $\beta_g=\infty$ we gauge-fix to an axial gauge by rotating all the $U$'s to
the identity. (Only $U=1$ configurations are important in the functional
integral in this limit.) Now $-{\cal L} = \beta_h \phi^\dagger(x) \phi(x+\mu)
$ which is a pure spin model. Again, one must analyze it on its own.

\begin{figure}
\centerline{\ewxy{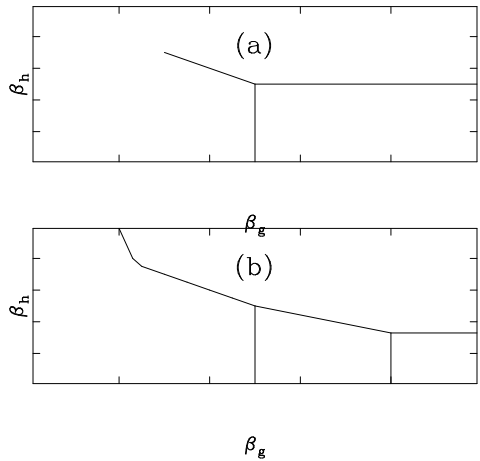}{80mm}
}
\caption{ Phase diagrams for (a)  $Z_2$ gauge-$Z_2$ spin and
and
(b) $Z_6$ gauge-$Z_3$ spin models.}
\label{fig:higgspd}
\end{figure}

Finally, we have $\beta_h \rightarrow \infty$. Use unitary gauge again.
In this limit $\phi_0^\dagger U_\mu(x)\phi_0$ is forced to its maximum value.
If it happens the $\phi$ transforms under a faithful 
representation of the gauge group G, only $U=1$ is allowed. If the representation of $\phi$ is not faithful, then there is a subgroup H of G in
which $U=1$. Along this line, we then have a gauge theory with gauge group
$H \in G$. In the first case the Higgs and confinement phases are likely
to be connected  \cite{OSFS} (as shown in Fig. \ref{fig:higgspd}a).
In the second case, the gauge symmetry on the top line is typically
a discrete symmetry, there is typically a first order phase transition 
(which extends into the phase diagram) and one must resort to Monte
Carlo to see if the phases are connected or separated.
Schematic phase diagrams for $Z_2$ gauge-$Z_2$ spin and
$Z_6$ gauge-$Z_3$ spin models  \cite{CREUZ} are shown in Figs. \ref{fig:higgspd}a and b.

Finally, a few words about fermions
(see the next section for details).
It is easy to study vectorlike 
theories on the lattice, and easiest to study multiples
of 2 or 4 degenerate flavors of fermions. It is reasonably easy to
study the chiral properties of model theories, although taking the chiral limit
in a computer simulation is hard. Common games to play
are to measure $\langle \bar \psi \psi \rangle$ in a simulation
at finite fermion mass and to extrapolate to zero mass.
With Kogut-Susskind fermions, this operator is only multiplicatively
renormalized, and so one can see if it vanishes at zero quark 
mass or extrapolates to some finite limit. One can also compute the
pion mass and ask whether $m_\pi^2 \simeq m_q$.
Only fermion actions which are bilinear, $S = \bar \psi M \psi$
for any matrix $M$, can be simulated on the computer, but if
one has a four-fermion interaction, the Hubbard-Stratonovich
transformation
\bee
\exp{-(\bar \psi \psi)^2 } =
\int d\phi \exp -{(\bar \psi \psi \phi + \phi^2)}
\ee
allows four-fermion interaction models to be converted into Yukawa models
(said differently, four-fermion interaction models are equivalent to
Yukawa models \cite{HUNGARIAN}),
 which the computer (and maybe the physicist) knows
how to solve.

\section{Standard Methods for Spectroscopy}

\subsection{Relativistic Fermions on the Lattice}

Defining fermions on the lattice presents a  new problem: doubling. The naive
procedure of discretizing the continuum fermion action results in a lattice model
with many more low energy modes than one originally anticipated. 
 Let's illustrate this with free field theory.

The free Euclidean fermion action in the continuum (in four dimensions) is
\bee
S = \int d^4 x [ \bar \psi(x) \gamma_\mu \partial_\mu \psi(x) + m \bar \psi(x)
\psi(x)  ] .\label{2.23}
\ee
One obtains the so-called naive lattice formulation by replacing the derivatives
by symmetric differences: we explicitly introduce the lattice spacing $a$
in the denominator and
write
\bee
S_L^{naive} = \sum_{n,\mu} \bar \psi_n {\gamma_\mu \over {2a}}
(\psi_{n+\mu} - \psi_{n-\mu}) + m \sum_n \bar \psi_n \psi_n . \label{2.24}
\ee
The propagator is:
\bee
G(p) = (i \gamma_\mu \sin p_\mu  a + ma)^{-1} 
= {{-i \gamma_\mu \sin p_\mu a + ma}\over{\sum_\mu \sin^2 p_\mu a + m^2 a^2}} 
\label{2.25}
\ee
We identify  the physical spectrum through the poles in the
propagator, at $p_0=iE$:
\bee
\sinh^2 Ea = \sum_j \sin^2 p_j a + m^2a^2
\ee
The lowest energy solutions are the expected ones
 at $p= (0,0,0)$, $E \simeq \pm m$, but  there are 
other degenerate
ones, at $p = (\pi,0,0)$, $(0,\pi,0,)$, \dots $(\pi,\pi,\pi)$.
This is a model for sixteen light fermions, not one.

\noindent
(a) Wilson Fermions

There are two ways to deal with the doublers.  The first way is to alter
 the dispersion relation so that it has only one low energy solution.  The
other solutions are forced to $E \simeq 1/a$ and become very heavy as $a$
is taken to zero.  The simplest version of this solution (and almost
the only one seen in the literature until recently) 
is due to Wilson: add a second-derivative-like term
\bee
S^W = -{r \over {2a}}\sum_{n,\mu}\bar \psi_n(\psi_{n+\mu} -2 \psi_n
+\psi_{n-\mu} ) \label{2.26}
\ee
to $S^{naive}$.  The parameter $r$ must lie between 0 and 1; $r=1$ is 
almost always used and  ``$r=1$'' is implied when one
speaks of using ``Wilson fermions.''  The propagator is
\bee
G(p) = {{-i \gamma_\mu \sin p_\mu a + m a -r \sum_\mu (\cos p_\mu a -1)} \over
{\sum_\mu \sin^2 p_\mu a + (m  a-r \sum_\mu(\cos p_\mu a -1))^2}} .
\label{2.27}
\ee
It has one pair of poles
 at $p_\mu \simeq (\pm im,0,0,0)$, plus other poles at $p \simeq r/a$.
In the continuum these states become infinitely massive and decouple
(although decoupling is not  trivial to prove).

With Wilson fermions it is conventional not to use not the mass but the
``hopping parameter'' $\kappa = {1 \over 2}(m a + 4r)^{-1}$, and to
rescale the fields $\psi \rightarrow \sqrt{2 \kappa} \psi$.  The action for an
interacting theory is conventionally written
\bee
S = \sum_n \bar \psi_n \psi_n -  \kappa \sum_{n \mu}(\bar \psi_n
(r - \gamma_\mu) U_\mu(n) \psi_{n+ \mu} + 
\bar \psi_n(r + \gamma_\mu) U_\mu^\dagger \psi_{n - \mu} ). \label{2.28}
\ee
Wilson fermions are closest to the continuum 
formulation-- there is a four component spinor on every lattice site for
every color and/or flavor of quark.  Constructing currents and states
is  just like in the continuum. 

However, the Wilson term explicitly breaks chiral symmetry.  
This has the 
consequence that the zero bare
 quark mass limit is not respected by interactions;
the quark mass is additively renormalized.    The value of $\kappa_c$,
the value of the  (bare)
hopping parameter at which the pion mass vanishes, is
not known a priori before beginning a simulation; it must be computed.
This is done in a simulation involving Wilson fermions 
 by varying $\kappa$ and watching  the pion mass 
extrapolate quadratically to zero as
$m_\pi^2  \simeq \kappa_c - \kappa$ ($\kappa_c - \kappa $ is proportional
to the quark mass for small $m_q$).

\noindent
(b) Staggered or Kogut-Susskind Fermions

In this formulation one reduces the number of fermion flavors by using
one component ``staggered'' fermion fields rather than four component Dirac
spinors.  The Dirac spinors are constructed by combining staggered fields
on different lattice sites.

Let us call the staggered field on site $n$ $\chi_n$ and define the matrix field\bee
\psi_{\alpha k}(n) = \sum_b (\gamma^{n+b})_{\alpha k} \chi_{n+b} \label{2.29}
\ee
where
\bee
\gamma^{n+b} = \gamma_1^{n_1 + b_1}
 \gamma_2^{n_2 + b_2}
 \gamma_3^{n_3 + b_3}
 \gamma_4^{n_4 + b_4} \label{2.30}
\ee
and $b$ runs over the sites of a hypercube
$b= (0,0,0,0)$, (1,0,0,0), \dots (1,1,1,1). $\bar \psi_{\alpha k}$ is
defined analogously,
\bee
\bar \psi_{\alpha k}(n) = \sum_b (\gamma^{n+b})^\dagger_{\alpha k}
 \bar \chi_{n+b} .\label{2.31}
\ee

One interprets the $\alpha$ indices as spin (= 1 to 4) and $k$ as flavor
(= 1 to 4).  The free action (with $m_q = 0$) is
\bee
S = {1 \over 128}\sum \bar \psi_n \gamma_\mu (\psi_{n+\mu} -\psi_{n - \mu}) \label{2.32}
\ee
\bee
 = {1 \over 2} \sum \bar \chi_n \eta_{n,\mu} (\chi_{n - \mu} - \chi_{n+\mu})
\label{2.33}
\ee
where $\eta_{n, \mu} = (-1)^{n_1+n_2+n_3+n_4}$.  Gauge fields can be
added in the standard way, and explicit mass terms give an extra
 $m_q \sum \bar \chi \chi$ term.

Staggered fermions preserve an explicit chiral symmetry as $m_q \rightarrow 0$
even for finite lattice spacing, as long as all four flavors are degenerate.
They are preferred over Wilson fermions in situations in which
the chiral properties of the fermions dominate the dynamics--for
example, in studying the chiral restoration/deconfinement transition
at high temperature.  They also present a computationally less intense
situation from the point of view of numerics than Wilson fermions, for the
trivial reason that there are less variables.
  However, flavor symmetry and translational
symmetry are all mixed together.  Construction of meson and baryon states
(especially the $\Delta$) is more complicated than for Wilson 
fermions \cite{GANDSMIT}.

\subsection{Enter the Computer}
A ``generic'' Monte Carlo simulation in QCD
breaks up naturally into two parts. In the ``configuration generation''
phase one constructs an ensemble of states with the appropriate
Boltzmann weighting: we compute observables simply by
averaging $N$ measurements using the field
variables $\phi^{(i)}$ appropriate to the sample
\bee
\langle{\Gamma}\rangle \simeq \bar \Gamma
\equiv {1 \over N}\sum_{i=1}^N\Gamma[\phi^{(i)}] .
\label{SAMPLE}
\ee
As the number of measurements $N$ becomes large the quantity $\bar \Gamma$
will become a  Gaussian distribution about a mean value.  Its standard
deviation is roughly   \cite{GPLTASI}
\bee
\sigma^2_\Gamma = {1 \over N}({1 \over N}\sum_{i=1}^N|\Gamma[\phi^{(i)}]|^2 -
\bar \Gamma^2). 
\label{STDEV}
\ee
The idea  of essentially all simulation algorithms
is that one constructs a new configuration
of field variables from an old one.  One begins with some simple field
configuration and monitor observables while the algorithm steps
along. After some number of steps, the value of observables will appear
to become independent of the starting configuration. At that point the
system is said to be ``in equilibrium'' and Eq. \ref{SAMPLE} can be
used to make measurements.

The simplest method for generating configurations is called the 
Metropolis  \cite{METROPOLIS}
algorithm. It works as follows:
From the old configuration $\{\phi\}$ with action $\beta S$, transform
the variables (in some reversible way) to a new trial configuration
$\{\phi\}'$ and compute the new action $\beta S'$. Then, if
$S' < S$ make the change and update all the variables; if not,
make the change with probability $\exp(-\beta(S'-S))$.

Why does it work? In equilibrium, the rate at which configurations $i$
turn into configurations $j$ is the same as the rate for the
back reaction $j \rightarrow i$. The rate of change is 
(number of configurations) $\times$ (probability of change). Assume
for the sake of the argument that $S_i < S_j$. Then the
rate $i \rightarrow j$ is $N_i P(i\rightarrow j)$ with
$P(i\rightarrow j)= \exp(-\beta(S_j-S_i)$ and  the
rate $j \rightarrow i$ is $N_j P(j\rightarrow i)$ with
$P(j\rightarrow i)= 1$. Thus $N_i/N_j = \exp(-\beta(S_i-S_j))$.

If you have any interest at all in the techniques I am describing, you should
write a little Monte Carlo program simulating the two-dimensional
Ising model. Incidentally, the
 favorite modern method for pure gauge models is overrelaxation  \cite{OVERRELAX}.

One complication for QCD which spin models don't have is
fermions. The fermion path integral is not a number and a computer can't
simulate fermions directly.  However, one can formally integrate out the
fermion fields. For $n_f$ degenerate flavors of staggered fermions
\bee
Z = \int [dU][d\psi][d\bar\psi] \exp(-\beta S(U) - \sum_{i=1}^{n_f}
\bar \psi M \psi)
\ee
\bee
=\int [dU](\det M)^{n_f/2}\exp(-\beta S(U)) .
\ee
(One can make the determinant positive-definite by
writing it as $\det(M^\dagger M)^{n_f/4}$.)
The determinant introduces a nonlocal interaction among the $U$'s:
\bee
Z = \int [dU] \exp(-\beta S(U)
 - {n_f \over 4} {\rm Tr} \ln (M^\dagger M)
 ) .  
\ee

All large scale dynamical fermion simulations today generate
configurations using some variation of the microcanonical ensemble. That is,
they introduce momentum variables $P$ conjugate to the $U$'s and integrate
Hamilton's equations through a simulation time $t$
\bee
\dot U = i P U 
\ee
\bee
\dot P = -{{\partial S_{eff}}\over{\partial U}} . 
\label{ALLAT}
\ee
The integration is done numerically by introducing a timestep $\Delta t$.
The momenta are repeatedly refreshed by bringing them in contact with
 a heat bath and the method is thus called Refreshed or Hybrid Molecular
Dynamics  \cite{HMD}.

For special values of $n_f$ (multiples of 2 for Wilson fermions or
 of 4 for staggered fermions) the equations of motion can be derived
from a local Hamiltonian and in that case $\Delta t$ systematics
in the integration can be removed by an extra Metropolis accept/reject
step.  This method is called Hybrid Monte Carlo  \cite{HMC}.
 
The reason for the use of these small timestep algorithms is that for
any change in any of the $U$'s, $(M^\dagger M)^{-1}$ must be recomputed.
When Eqn. \ref{ALLAT} is integrated all of the $U$'s in the lattice
are updated simultaneously, and only one matrix inversion is needed per
change of all the bosonic variables.
 
The major computational problem dynamical fermion simulations face is inverting
the fermion matrix $M$. It has eigenvalues with a very large range--
from $2\pi$ down to $m_q a$-- and in the physically interesting limit of
small $m_q$ the matrix becomes ill-conditioned.  At present it is necessary
to compute at unphysically heavy values of the quark mass and to extrapolate
to $m_q=0$.  The standard inversion technique today is one of the variants of
the  conjugate gradient algorithm \cite{FROMMER}.

\subsection{Spectroscopy Calculations}

``In a valley something like a race took place. A little crowd watched
bunches of cars, each consisting of two `ups' and a `down' one, starting
in regular intervals and disappearing in about the same direction. `It
is the measurement of the proton mass,' commented Mr. Strange, `they
have done it for ages. A very dull job, I am glad I am not in the game.' ''
 \cite{SHURYAK}

Masses are computed in lattice  simulations from the
asymptotic behavior of Euclidean-time
 correlation functions.  A typical (diagonal) correlator can be written as
\bee
C(t) = \langle 0 | O(t) O(0) | 0\rangle  .  
\ee
Making the replacement
\bee
O(t)=e^{Ht}Oe^{-Ht} 
\ee
and inserting a complete set of energy eigenstates,  Eq. (3.1) \ becomes
\bee
C(t) = \sum_n |\langle 0 |  O|n\rangle |^2 e^{-E_nt}.  
\ee
At large separation the correlation function is approximately
\bee
C(t) \simeq  |\langle 0 |  O|1\rangle |^2 e^{-E_1t} 
\label{CORRFN}
\ee
where $E_1$ is the energy of the lightest state which the operator $O$
can create from the vacuum. 
If the operator does not couple to the vacuum, then
in the limit  of large $t$ one hopes to to find
the mass $E_1$
by measuring the leading exponential falloff of the correlation function,
and most lattice simulations begin with that measurement.
If the operator $O$ has poor overlap with the lightest state,
a reliable value for the mass can be extracted only at a large time $t$.
In some cases that state is the vacuum itself,
in which $E_1 = 0$.  
Then one looks for the next higher state--a signal which disappears
into the constant background.
This makes the actual calculation of the energy  more difficult.
 
This is the basic way hadronic masses are found in lattice gauge theory.  The
many calculations differ in important specific details of choosing the
 operators
$O(t)$.

\subsection{Recent Results}

 Today's supercomputer
QCD simulations range from $16^3 \times 32$ to
 $32^3 \times 100$
points and run from hundreds (quenched) to thousands (full QCD) of
hours on the fastest supercomputers in the world. 

Results are presented in four common ways. Often one sees a plot
of some bare parameter vs. another bare parameter. This is not
very useful if one wants to see continuum physics, but it is how
we always begin. Next, one can plot a dimensionless
ratio as a function of the lattice spacing. These plots
represent quantities like Eqn. \ref{SCALING}. Both axes can show
 mass ratios.
Examples of such plots  are the so-called
Edinburgh plot ($m_N/m_\rho$ vs. $m_\pi/m_\rho$) and the Rome plot
 ($m_N/m_\rho$ vs. $(m_\pi/m_\rho)^2$).   These plots can answer 
continuum questions
(how does the nucleon mass change if  the quark mass is changed?)
or can be used to show (or hide) scaling violations.
Plots of one quantity in MeV vs. another quantity in MeV are typically
rather heavily processed after the data comes off the computer.

Let's look at some examples of spectroscopy, done in the ``standard way,''
with the plaquette gauge action and Wilson or staggered quarks.
I will restrict the discussion to quenched simulations because
only there are the statistical errors small enough to be interesting to
a non-lattice audience.
Most dynamical fermion simulations are unfortunately like Dr. Johnson's
dog.

\begin{figure}
\centerline{\ewxy{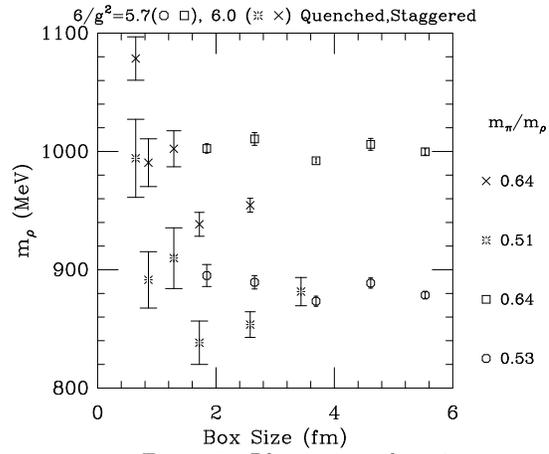}{80mm}
}
\caption{ Rho mass vs. box size.}
\label{fig:mrhovsnsa}
\end{figure}

Fig. \ref{fig:mrhovsnsa} shows a plot of the rho mass as a function
of the size of the simulation, for several values of the quark mass
(or $m_\pi/m_\rho$ ratio in the simulation) and lattice spacing
($\beta=6.0$ is $a \simeq 0.1$ fm and $\beta=5.7$ is about twice that)
 \cite{MILC}.
This picture shows that if the box has a diameter bigger than about 2 fm,
the rho mass is little affected, but if the box is made smaller, the rho is ``squeezed'' and its mass rises.

\begin{figure}
\centerline{\ewxy{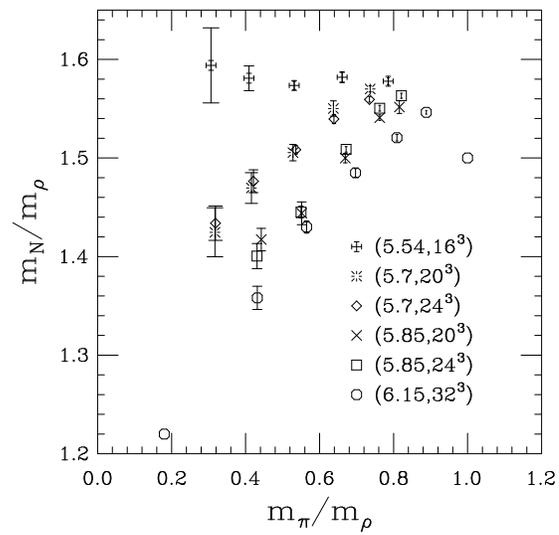}{80mm}
}
\caption{ An Edinburgh plot for staggered fermions, 
from the MILC collaboration.}
\label{fig:edinburgh}
\end{figure}

\begin{figure}
\centerline{\ewxy{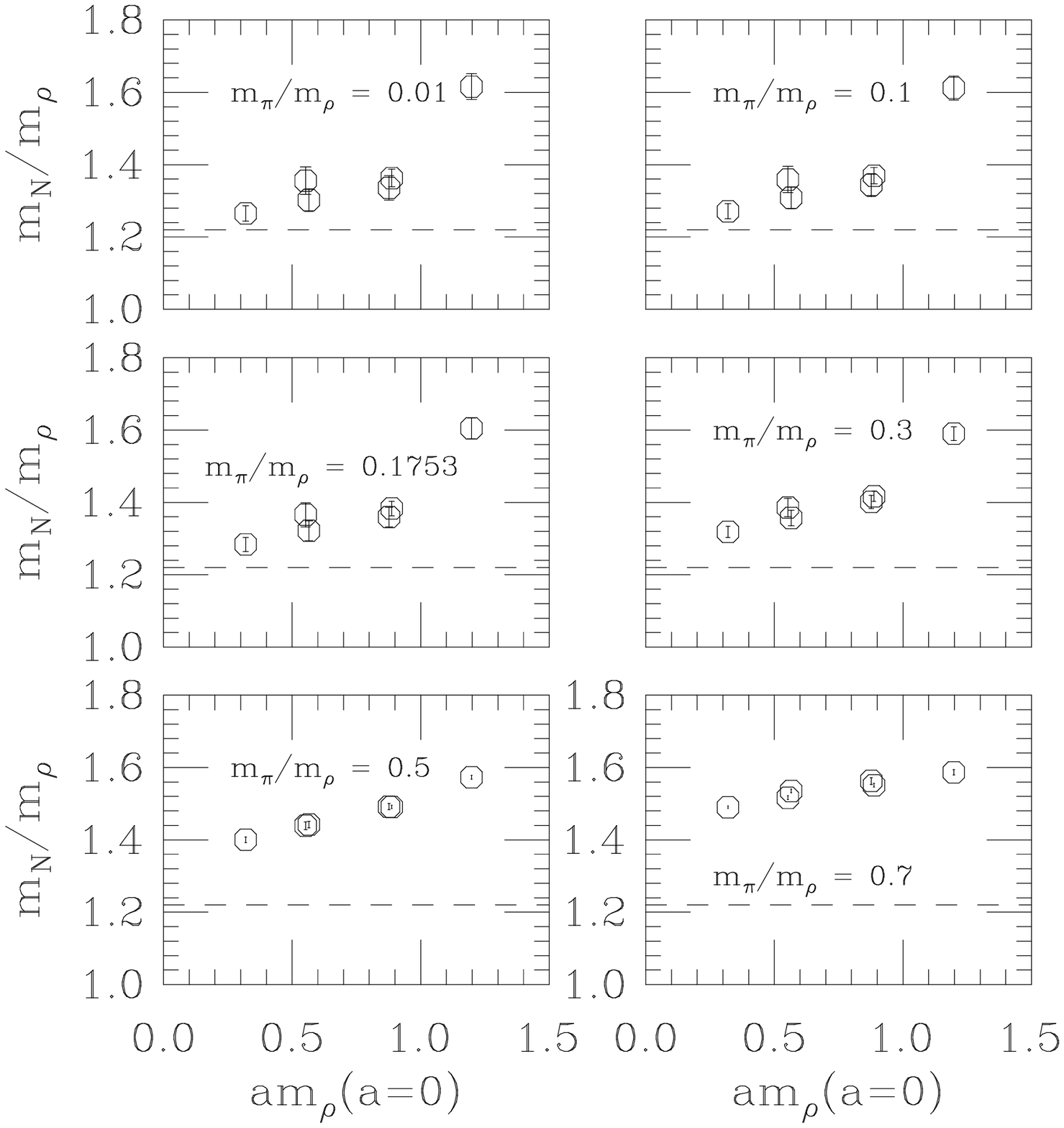}{80mm}
}
\caption{ ``Sections'' through the Edinburgh plot.}
\label{fig:fig3combo}
\end{figure}

Next we look at an Edinburgh plot, Fig. \ref{fig:edinburgh}  \cite{MILC}.
The different plotting symbols correspond to different bare couplings
or (equivalently) different lattice spacings.
This plot shows large scaling violations: mass ratios from different
lattice spacings do not lie on top of each other. We can expose the level of
scaling violations by taking ``sections'' through the plot and plot
$m_N/m_\rho$ at fixed values of the quark mass (fixed $m_\pi/m_\rho$), 
vs. lattice spacing, in
Fig. \ref{fig:fig3combo}.

\begin{figure}
\centerline{\ewxy{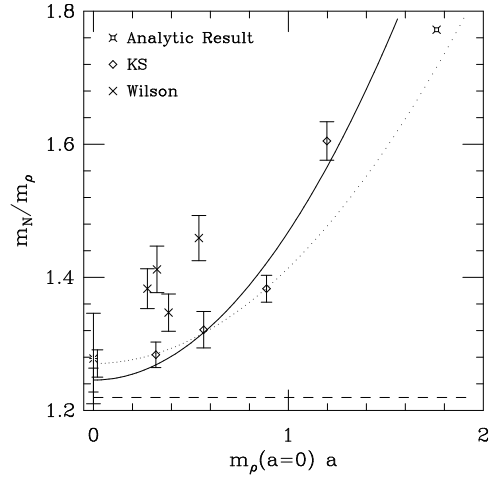}{80mm}
}
\caption{ Nucleon to rho mass ratio (at chiral limit) vs. lattice spacing
(in units of $1/m_\rho$).}
\label{fig:ratiovsmrhoa}
\end{figure}

Now for some examples of scaling tests in the chiral limit.
(Extrapolating to the chiral limit is a whole can of worms on its own,
but for now let's assume we can do it.)
Fig. \ref{fig:ratiovsmrhoa} shows the nucleon
to rho mass ratio (at chiral limit) vs. lattice spacing
(in units of $1/m_\rho$) for staggered  \cite{MILC} and Wilson   \cite{IBM}
fermions.
The ``analytic'' result is from strong coupling. The two curves
are quadratic extrapolations to zero lattice spacing using different
sets of points from the staggered data set. The burst is from a linear
extrapolation to the Wilson data. The reason I show this figure
is that one would like to know if the continuum limit of
quenched spectroscopy ``predicts'' the real-world $N/\rho$ mass ratio
of 1.22 or not. The answer (unfortunately) depends on how the reader
chooses to extrapolate.

\begin{figure}
\centerline{\ewxy{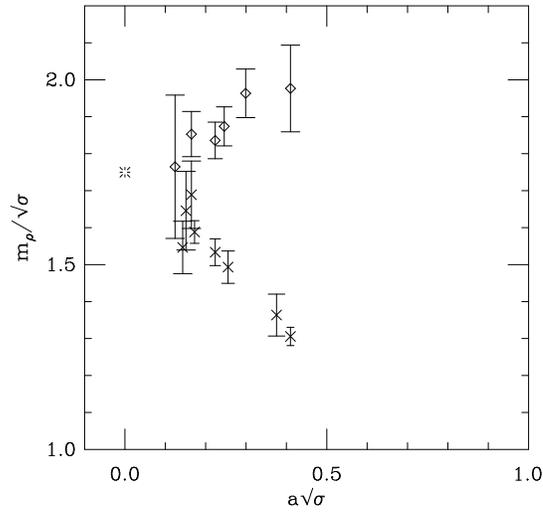}{80mm}
}
%
\caption{Scaling test for the rho mass in terms of the string tension,
with data points labeled as in Fig. 8.}
\label{fig:sommerfig}
\end{figure}

Another test  \cite{SOMMERPLOT}
 is the ratio of the rho mass to the square root of
the string tension, Fig. \ref{fig:sommerfig}.
 Here the diamonds are staggered data and the crosses
from the Wilson action. Scaling violations are large but the eye
extrapolates to something close to data (the burst).

\begin{figure}
\centerline{\ewxy{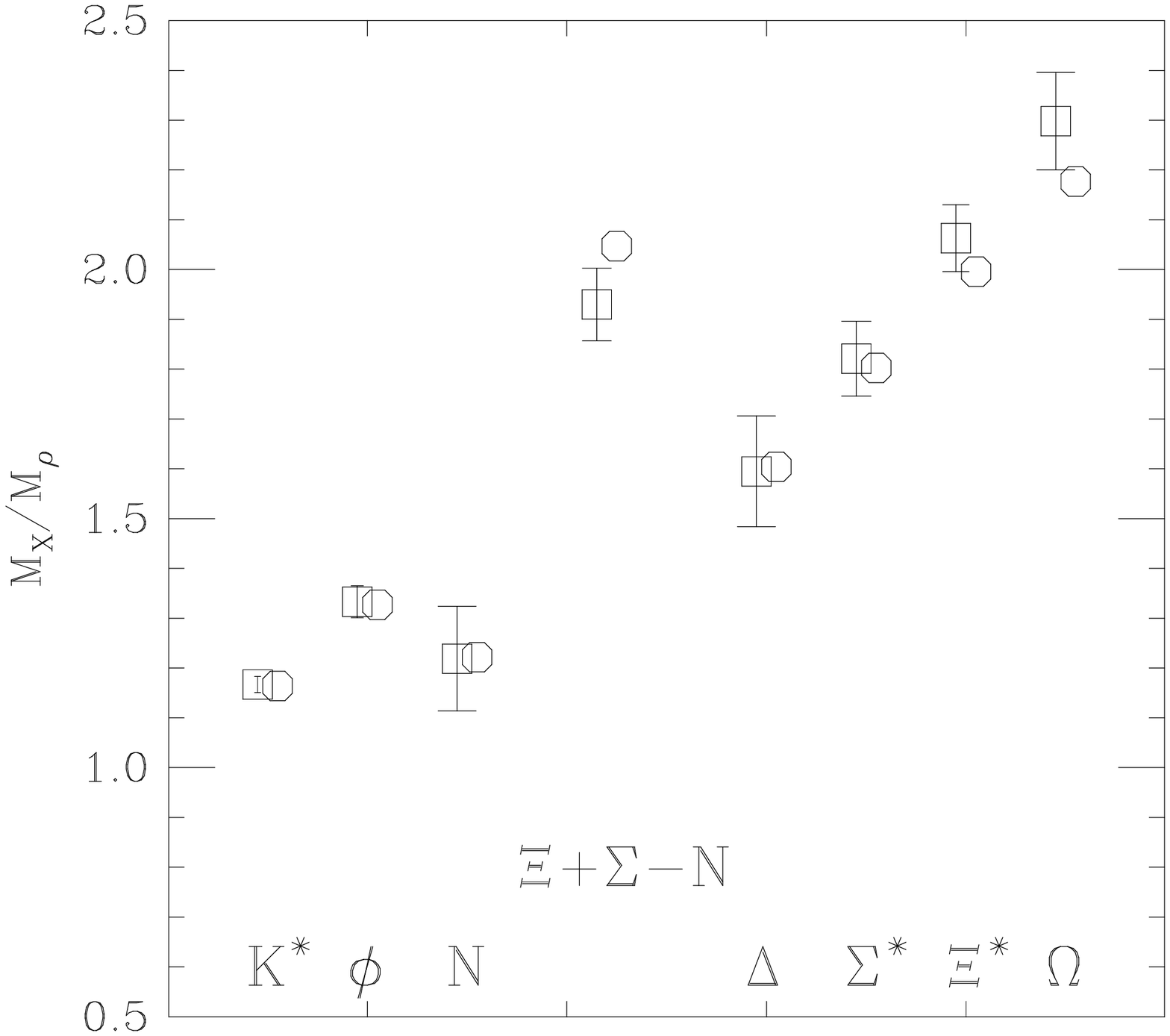}{80mm}
}
\caption{Quenched approximation mass ratios from Ref. 25.}
\label{fig:ratios}
\end{figure}

Finally, despite Mr. Strange, very few authors have attempted to
extrapolate to infinite volume, zero lattice spacing, and to physical quark
masses, including the strange quark. One group which did, 
Butler et al.  \cite{IBM}, produced Fig. \ref{fig:ratios}. The
squares are lattice data, the octagons are the real world. They look
quite similar within errors. Unfortunately, to produce this picture, they had
to build their own computer.

\section{Bare Actions, Scaling, and Improved Actions}

(Disclaimer: It happens that at present
there are a number of approaches to ``improvement.''
At present, the words which are used to describe them differ greatly
among the different groups, and the physics motivations are quite different.
For example, in one approach, ideas based on perturbation theory
are regarded as absolutely crucial while in another approach
perturbation theory is said to be completely irrelevant.
The particular problems whose solution would be regarded
as a success are different for different groups. This makes life
exciting for the researcher but dangerous for the reviewer.
You should know at the start that I work in this area, and
that I am as prejudiced as anyone else in the field.)

The slow approach to scaling presents a practical problem for QCD
simulations, since it means that one needs to work at small lattice
spacing. This is expensive. The cost of a Monte Carlo simulation
in a box of physical size $L$ with lattice spacing $a$ and quark mass
$m_q$ scales roughly as
\bee
({L \over a})^4 ({1\over a})^{1-2}({1 \over m_q})^{2-3}
\label{COST}
\ee
where the 4 is just the number of sites, the 1-2 is the cost of
``critical slowing down''--the extent to which successive configurations
are correlated, and the 2-3 is the cost of inverting the fermion
propagator, plus critical slowing down from the nearly massless
pions. 
The problem  is that one needs a big computer to do anything.

However, all the simulations I described in the last lecture were
done with a particular choice of lattice action: the plaquette
gauge action, and either Wilson or staggered quarks. While those actions are
the simplest ones to program, they are just particular arbitrary choices of
bare actions. Can one invent a better lattice discretization, which
has smaller scaling violations?

To approach this problem, let's think about the connection
between scaling and the properties of some arbitrary bare action,
which we assume is defined with some UV cutoff $a$ (which does not
have to be a lattice cutoff, in principle). The action is
characterized by an infinite number of coupling constants, $\{g\}$.
When the $g$'s have any arbitrary value, the typical scale for all
physics will be the order of the cutoff: $m \simeq 1/a$, correlation
length $\xi \simeq a$. There will be strong cutoff effects.

The best way to think about scaling is through the renormalization
group  \cite{WK}. Take the action with cutoff $a$ and integrate out degrees
of freedom to construct a new effective action with a new cutoff
$a' > a$ and fewer degrees of freedom. The physics at distance scales
$r > a$ is unaffected by the coarse-graining (assuming it is done
exactly.) We can think of the effective actions as being similar to the
original action, but with altered couplings. We can repeat this
coarse-graining and generate new actions with new cutoffs.
As we do, the coupling constants ``flow:''
\bee
S(a,c_j) \rightarrow S(2a,c_j') \rightarrow S(4a,c_j'')\rightarrow \dots
\ee
If under repeated blockings the system flows to a fixed point
\bee
S(a_n,c^n_j) \rightarrow S(a_{n+1},c_j^{n+1}=c_j^n)
\ee
then observables are independent of the cutoff $a$ and in particular
the correlation length  $\xi$ must either be zero or infinite.

This can only happen if the original $c$'s belong to a particular
 restricted set,
called the ``critical surface.'' It is easy to see that physics
on the critical surface is universal: at long distances the effective
 theory is the action at the fixed point, to which all the couplings 
have flowed,
regardless of their original bare values.

But $\xi = \infty$ is not $\xi$ large. Imagine tuning bare parameters 
close to the critical surface, but not on it. The system will 
flow towards the fixed point, then away from it.  The flow lines in coupling
constant space will asymptotically approach a particular trajectory,
called the renormalized trajectory, which connects (at $\xi = \infty$)
with the fixed point. Along the renormalized trajectory, $\xi$ is finite.
However, because it is connected to the fixed point, it shares 
the scaling properties of the fixed point--in particular, the 
complete absence of cutoff effects in the spectrum and in Green's functions.
(To see this remarkable result, imagine doing  QCD
spectrum calculations with the original bare action with a cutoff
equal to the Planck mass-or beyond (!) and then coarse graining.
Now exchange the order of the two procedures.
 If this can be done without making any approximations the
answer should be the same.)

\begin{figure}
\centerline{\ewxy{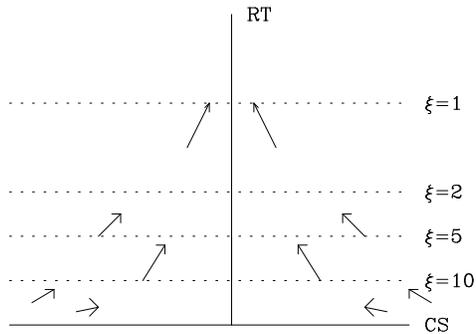}{80mm}
}
\caption{ A schematic picture of renormalization group flows along a one-dimensional critical surface, with the associated renormalized
trajectory, and superimposed contours of constant correlation length.}
\label{fig:rt}
\end{figure}

A Colorado analogy is useful: think of the critical surface as the top
of a high mountain ridge. The fixed point is a saddle point on the ridge.
A stone released on the ridge will roll to the saddle and come to rest.
 If it is not released exactly on the ridge, it will roll near
to the saddle, then go down the gully leading away from it.
For a cartoon, see Fig. \ref{fig:rt}.

So the ultimate goal of ``improvement programs'' is to find a true
perfect action, without cutoff effects, along the renormalized trajectory
of some renormalization group transformation.

The bare action for an asymptotically free theory like QCD can be
characterized by one weakly relevant operator, $F_{\mu\nu}^2$, and 
its associated coupling $g^2$, plus many irrelevant operators. In lattice
language, our bare actions are described by one overall factor of 
$\beta=2N/g^2$
and arbitrary weights of various closed loops,
\bee
\beta S = {{2N} \over g^2} \sum_j c_j O_j.
\ee
Asymptotic freedom is equivalent to the statement that
the critical surface of any renormalization group transformation
is at $g^2=0$. The location of a fixed point
involves some relation among the $c_j$'s.

As you might imagine, finding a renormalized trajectory is hard, and nobody
has done it in a convincing way yet for QCD. What are people doing
in the meantime?

\subsection{Improvement based on naive dimensional analysis}

The idea here is that since the critical surface
 is at $g^2\rightarrow 0$ we can
use the naive canonical dimensionality of operators
 to guide us in our choice of improvement.
If we perform a naive Taylor expansion of a lattice operator like
the plaquette, we find that it can be written as
\beea
1 - {1 \over 3} {\rm Re \ Tr} U_{plaq} = &
                  r_0 {\rm Tr} F_{\mu\nu}^2  
+a^2 [ r_1 \sum_{\mu\nu}{\rm Tr} D_\mu F_{\mu\nu} D_\mu F_{\mu\nu} +
\nonumber \\
 &  r_2 \sum_{\mu\nu\sigma}{\rm Tr} D_\mu 
F_{\nu\sigma} D_\mu F_{\nu\sigma} + \nonumber \\
 &  r_3 \sum_{\mu\nu\sigma}{\rm Tr} D_\mu F_{\mu\sigma} D_\nu F_{\nu\sigma}]+
 \nonumber \\
 & +O(a^4) 
\eea
The expansion coefficients have a power series expansion in the
coupling, $r_j = A_j + g^2 B_j + \dots$
and the expectation value of any operator $T$ will have an expansion
\bee
\langle T(a) \rangle = \langle T(0) \rangle + O(a) + O(g^2 a) + \dots
\ee

Other loops have a similar expansion, with different coefficients.
Now the idea is to take a minimal subset of loops and systematically remove the
 $a^n$ terms order by order
for physical observables by taking the right linear combination of
loops. 
\bee
S = \sum_j c_j O_j
\label{EXPA1}
\ee
with
\bee c_j = c_j^0 + g^2 c_j^1 + \dots
\label{EXPA2}
\ee 
``Tree level'' improvement removes pure power-law corrections through
some order in $a^n$.
One can also consider quantum corrections,
and at each order in $a^n$ remove $g^m$ terms, too.
This method was developed by Symanzik and 
co-workers  \cite{SYMANZIK}  \cite{WEISZ}  \cite{LWPURE} ten years ago.

\begin{figure}
\centerline{\ewxy{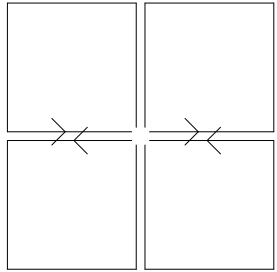}{80mm}
}
\caption{ The ``clover term''.}
\label{fig:clover}
\end{figure}

The most commonly used ``improved'' fermion action
is the
``Sheikholeslami-Wohlert''  \cite{SHWO} or ``clover'' action, an order $a^2$
improved Wilson action. The original
Wilson action has $O(a)$ errors in its vertices, $S_W = S_c + O(a)$.
This is corrected by making a field redefinition
\beea
\psi(x) \rightarrow \psi'(x) = & \psi(x) + {{ia}\over 4} \gamma_\mu D_\mu\psi
\\
\bar\psi(x) \rightarrow \bar\psi'(x) = & \bar \psi(x) + {{ia}\over 4} \gamma_\mu \bar\psi D_\mu
\eea
and the net result is an action with an extra lattice anomalous magnetic
moment term,
\bee
S_{SW} - {{iag}\over 4} \bar \psi (x)\sigma_{\mu\nu}F_{\mu\nu} \psi(x)
\ee
It is called the ``clover'' action because the lattice version
of $F_{\mu\nu}$ is the sum of paths shown in Fig. \ref{fig:clover}.

Studies performed at the time showed that this program
 did not improve scaling for the pure gauge theory
(in the sense that the cost of simulating the more complicated action 
was greater than the savings from using a larger lattice spacing.)
The whole program was re-awakened in the last few last years
 by Lepage and collaborators  \cite{PETERIMP} and variations of this program
give the most widely used ``improved'' lattice actions.

\subsection{Nonperturbative determination of coefficients}
\begin{figure}
\centerline{\ewxy{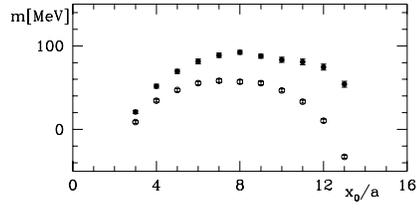}{80mm}
}
\caption{ Values of the quark mass as computed from the axial
and pseudoscalar currents, using the Wilson action. The open
and full symbols correspond to different boundary conditions on the
gauge fields.}
\label{fig:before}
\end{figure}

\begin{figure}
\centerline{\ewxy{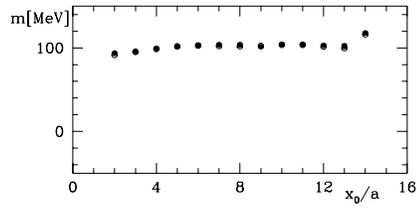}{80mm}
}
\caption{ Same as previous figure, but now with improved action
and operators.}
\label{fig:after}
\end{figure}

Although I am breaking chronological order, the simplest approach
to Symanzik improvement is the newest. The idea   \cite{NPIMP} is to force
the lattice to obey various desirable identities to some order in $a$,
by tuning parameters until the identities are satisfied by the simulations.
Then calculate other things. One example is the PCAC relation
\bee
\partial_\mu A_\mu^a = 2m_ P^a + O(a),
\ee
where the axial and pseudoscalar currents are just
\bee
A_\mu^a(x) = \bar \psi(x) \gamma_\mu\gamma_5 {1\over 2} \tau^a \psi(x)
\ee
and
\bee
P^a(x) = \bar \psi(x) \gamma_5 {1\over 2} \tau^a \psi(x)
\ee
($\tau^a$ is an isospin index.) The PCAC relation for the quark mass is
\bee
m \equiv  {1\over 2} {{\langle \partial_\mu A_\mu^a O^a \rangle} \over
{\langle P^a O^a \rangle}} + O(a).
\label{quarkmass}
\ee
Now the idea is to take some Symanzik-improved action, with
the expansion coefficients allowed to vary, 
and perform simulations in a little box with some particular choice of
boundary conditions for the fields. Tune the parameters, which include
the
 $i/4 c_{SW} a \sigma_{\mu\nu}F_{\mu\nu}$ and redefined  currents
\bee
A_\mu = Z_A[(1+b_A a m_q)A_\mu^a + c_A a \partial_\mu P^a
\ee
\bee
P^a = Z_P(1+ b_P a m_q)P^a,
\ee
until the quark mass, defined in Eq. \ref{quarkmass}, is independent
of location in the box, or of the boundary conditions.
Figs. \ref{fig:before} and \ref{fig:after} illustrate what can be done
with this tuning procedure.  It is still too soon for definitive tests
of scaling with this procedure.

\subsection{Improving perturbation theory}
The older version of Symanzik improvement uses lattice perturbation
theory to compute the coefficients of the operators in the action.
Let's make a digression into lattice perturbation theory \cite{MORNINGSTAR}.
It has three major uses.
First, we need to
relate lattice quantities (like matrix elements) to continuum ones:
$O^{cont}(\mu) = Z(\mu a, g(a))O^{latt}(a).$ 
This happens because the renormalization of an operator is slightly different
in the two schemes. In perturbation theory $Z$ has an expansion in powers of
$g^2$.
Second, we can use perturbation theory
to understand and check numerical calculations when the lattice 
couplings are very small.
Finally, one can use perturbative ideas to
 motivate nonperturbative improvement schemes \cite{PETERPAUL}.

Perturbation theory for lattice actions is just like any other
kind of perturbation theory (only much messier). One expands the
Lagrangian into a quadratic term and interaction terms
and constructs the propagator from the quadratic terms:
\beea
{\cal L} = & A_\mu(x) \rho_{\mu\nu}(x-y) A_\nu(y) + g A^3 + \dots \\
          =& {\cal L}_0 + {\cal L}_I .
\eea
For example, the gluon propagator in Feynman gauge for the Wilson
action is
\bee
D_{\mu\nu}(q) = {{g_{\mu\nu}}\over{ \sum_\mu(1-\cos(q_\mu a))}}.
\ee
To do perturbation theory for any system (not just the lattice)
one has to do three things:
one has to fix the renormalization scheme (RS) (define a coupling), 
specify the scale
at which the coupling is defined, and determine a numerical value for the 
coupling at that scale. All of these choices are arbitrary, and any perturbative
calculation is intrinsically ambiguous.

 Any object which has a perturbative expansion
can be written
\bee
O(Q) = c_0 + c_1(Q/\mu,RS) \alpha_s(\mu,RS) +c_2(Q/\mu,RS) \alpha_s(\mu,RS)^2
+\dots
\ee
In perturbative calculations we truncate the series after a fixed number
of terms and implicitly assume that's all there is. The  coefficients
$c_i(Q/\mu,RS)$ and the  coupling $\alpha_s(\mu,RS)$ depend on
the renormalization scheme  and choice of scale $\mu$.
 The guiding rule of perturbation theory \cite{MORNINGSTAR} is
``For a good choice of expansion the uncalculated higher order terms
should be small.''
 A bad choice has big coefficients.

There are many ways to define a coupling:
The most obvious is the bare coupling; as we will see shortly,
it is a poor expansion parameter. Another possibility is to define the
coupling from some physical observable. One popular choice is to use
the heavy quark potential at very high momentum transfer to define
\bee
V(q) \equiv 4 \pi C_f {{\alpha_V(q)}\over q^2}.
\ee
There are also several possibilities for picking a scale:
One can use the bare coupling, then $\mu=1/a$ the lattice spacing.
One can guess the scale or or play games just like in the continuum.
One game is the Lepage-Mackenzie $q^*$ prescription: take
\bee
\alpha_s(q^*)\int d^4 q \xi(q) = \int d^4 q \xi(q) \alpha_s(q).
\label{BLME}
\ee
To find $q^*$, write $\alpha_s(q) = \alpha_s(\mu) +
b \ln(q^2/\mu^2)\alpha_s(\mu)^2 +\dots$,
and similarly for $\alpha_s(q^*)$, insert these expressions
into Eq. \ref{BLME} and compare the
$\alpha_s(\mu)^2$ terms, to get
\bee
\ln(q^*) = \int d^4 q \ln(q) \xi / \int d^4 q \xi .
\label{IMMP}
\ee
This is the lattice analog of the Brodsky-Lepage-Mackenzie  \cite{BLM}
prescription in continuum PT.

\begin{figure}
\centerline{\ewxy{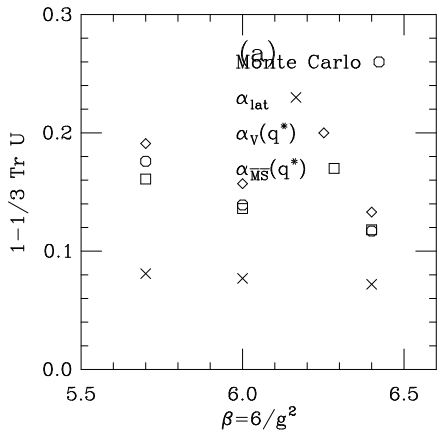}{80mm}
\ewxy{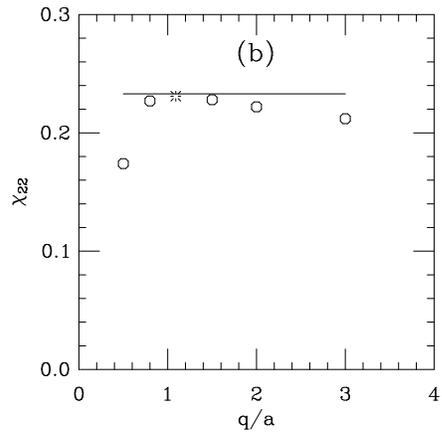}{80mm}}
\vspace{0.5cm}
\centerline{\ewxy{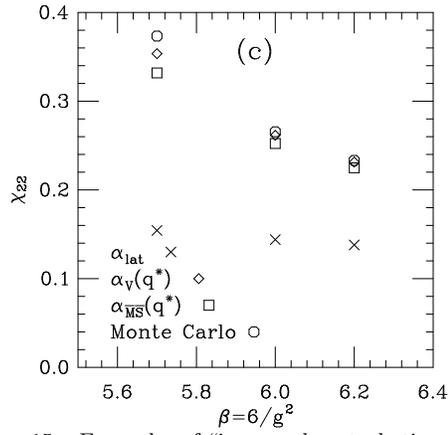}{80mm}
}
\caption{ Examples of ``improved perturbation theory''.}
\label{fig:improve}
\end{figure}

Finally one must determine the coupling:
If one uses the bare lattice coupling it is already known. Otherwise,
one can compute it in terms of the bare coupling:
\bee
\alpha_{\overline{MS}}(s/a) = \alpha_0 + (5.88-1.75 {\rm ln} s) \alpha_0^2
+ (43.41 - 21.89 {\rm ln} s + 3.06 {\rm ln}^2 s) \alpha_0^3 + \dots
\ee
Or one can determine it from something one measures on the lattice,
which has a perturbative expansion. For example
\bee
-{\rm ln}\langle {1\over 3} {\rm Tr}U_{plaq} \rangle=
{4\pi \over 3}\alpha_P(3.41/a)(1-1.185\alpha_P)
\label{PLAQALPHA}
\ee
(to this order, $\alpha_P=\alpha_V$).
Does ``improved perturbation theory'' actually improve
 perturbative calculations?
In many cases, yes: some examples are shown in Fig. \ref{fig:improve}
from  \cite{PETERPAUL}:
On the upper left we see a calculation of the average link in Landau
gauge, from simulations (octagons) and then from lowest-order
perturbative calculations using the bare coupling (crosses)
and $\alpha_V$ and $\alpha_{\bar {MS}}$ (diamonds and squares).
In the upper right panel we see how the lattice prediction of
 an observable involving the 2 by 2 Wilson loop depends on the
choice of momentum $q^*/a$ (at $\beta=6.2$, a rather weak value of the
bare coupling)
in the running coupling constant. The
burst is the value of the prescription of Eq. \ref{IMMP}.
In the lower panel are perturbative predictions the same observables
as a function of lattice coupling. These pictures illustrate
that perturbation theory in terms of the bare coupling does not work well,
but that using other definitions for
couplings, one can get much better agreement with
the lattice ``data''.

\begin{figure}
\centerline{\ewxy{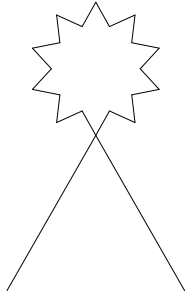}{80mm}
}
\caption{ The ``tadpole diagram''.}
\label{fig:tadpole}
\end{figure}

Straight perturbative expansions by themselves for the commonly-used lattice
actions  are typically not very convergent. The culprit is
the presence of $U_\mu$'s in the action. One might think that for weak
coupling, one could expand
\bee
\bar\psi U \psi = \bar \psi [ 1 + iga A + \dots ] \psi
\ee
and ignore the $\dots$, 
but the higher order term $\bar \psi {1\over 2}g^2a^2 A^2   \psi$
generates the ``tadpole graph'' of Fig. \ref{fig:tadpole}.
The UV divergence in the gluon loop $\simeq 1/a^2$ cancels the
$a^2$ in the vertex. The same thing happens at higher order,
and the tadpoles add up to an effective $a^0\sum c_ng^{2n}$ contribution.
Parisi  \cite{PARISI} and later
Lepage and Mackenzie  \cite{PETERPAUL} suggested a heuristic way
to deal with this problem: replace
$U_\mu \rightarrow u_0(1+igaA)$ where $u_0$, the ``mean field term'' or
``tadpole improvement term'' is introduced phenomenologically to sum the loops.
Then one rewrites the action as
\bee
S = {1 \over {g^2 u_0^4}} \sum {\rm Tr}U \leftrightarrow {1\over{g_{lat}^2}}
\sum {\rm Tr U}
\ee
where $g^2 \equiv g_{lat}^2/u_0^4$ is the new expansion parameter.
Is $u_0^4 \equiv \langle {\rm Tr}U_{plaq}/3\rangle$? This choice is often
used; it is by no means unique.

A ``standard action'' (for this year, anyway) is the
``tadpole-improved L\"uscher-Weisz  \cite{LWPURE}
 action,'' composed of a 1 by 1, 
1 by 2, and ``twisted'' loop (+x,+y,+z,-x,-y,-z),
\bee
\beta S = -\beta [ {\rm Tr}(1\times 1) - {1 \over {20 u_0^2}}
(1+0.48 \alpha_s){\rm Tr}(1 \times 2)
- {1 \over u_0^2} 0.33 \alpha_s {\rm Tr}U_{tw} ]
\ee
with $u_0 \equiv  \langle {\rm Tr}U_{plaq}/3\rangle^{1/4}$
and $3.068 \alpha_s \equiv -\ln\langle {\rm Tr}U_{plaq}/3\rangle$
determined self-consistently in the simulation.

None of this section seems to have anything to do with critical
surfaces or renormalized trajectories. What is going on?
I believe that the connection is that people following this path
to improvement are (implicitly)
trying to find a trajectory in coupling constant space
for which physical observables have no cutoff effects through
some order in $a^n$ and/or $(g^2)^m a^n$. 
The trajectory is described by the variation of the
$c_j$'s in Eqn. \ref{EXPA1} with respect to $a$ or to bare $g^2$.
If perturbation theory is reliable, one could compute them using
the expansion of Eqn. \ref{EXPA2}.
This is not a renormalized trajectory, but it might be
``good enough'' in the engineering sense.

\subsection{Fixed Point Actions}
A more direct attack on the renormalized trajectory begins by
finding a fixed point action. Imagine having a set of
field variables $\{\phi\}$ defined with a cutoff $a$. Introduce
some coarse-grained variables $\{\Phi\}$ defined with respect to
a new cutoff $a'$, and integrate out the fine-grained variables to
produce a new action
\bee
\rme^{-\beta S'(\Phi)} = \int d\phi  \rme^{-\beta(T(\Phi,\phi)+S(\phi))}
\label{RGE}
\ee
where $\beta(T(\Phi,\phi)$ is the blocking kernel which
functionally relates the coarse and fine variables. Integrating
Eq. \ref{RGE} is usually horribly complicated. However, P. Hasenfratz
and F. Niedermayer   \cite{HN} noticed an amazing simplification for
asymptotically free theories: Their critical surface is at $\beta=\infty$
and in that limit Eq. \ref{RGE} becomes a steepest-descent relation
\bee
S'(\Phi) = \min_{\phi}((T(\Phi,\phi)+S(\phi))
\ee
which can be used to find the fixed point action
\bee
S_{FP}(\Phi) = \min_{\phi}((T(\Phi,\phi)+S_{FP}(\phi)).
\label{RGFP}
\ee
The program has been successfully carried out for $d=2$ 
sigma models  \cite{HN} and for four-dimensional pure gauge theories
 \cite{PAPER123}.  These actions have two noteworthy properties:
First, not only are they classically perfect actions (they have no
$a^n$ scaling violations for any $n$), but they are also one-loop
quantum perfect: that is, as one moves out the renormalized trajectory,
\bee
{1\over g^2} S_{RT}(g^2) = {1 \over g^2}(S_{FP} + O(g^4) ).
\label{QPF}
\ee
Physically this happens because the original action has no irrelevant
operators, and they are only generated through loop graphs.
Thus these actions are an extreme realization of the Symanzik
program. Second, because these actions are at the fixed point, they
have scale invariant classical solutions. This fact can be used
to define a topological charge operator on the lattice in
a way which is consistent with the lattice action  \cite{INSTANTON12}.

These actions are ``engineered'' in the following way: one picks
a favorite blocking kernel, which has some free parameters, and
solves Eq. \ref{RGFP}, usually approximately at first.
Then one tunes the parameters in the kernel to optimize the 
action for locality, and perhaps refines the solution.
Now the action is used in simulations at finite correlation
length (i.e. do simulations with a Boltzman factor
$\exp(-\beta S_{FP})$. Because of Eq. \ref{QPF}, one believes
that the FP action will be a good approximation to the
perfect action on the RT; of course, only a numerical test can tell.
As we will see in the next section, these actions perform very well.
At this point in time no nonperturbative
FP action which includes fermions has been tested, but most of
the formalism is there  \cite{WIESE}.

\subsection{Examples of ``Improved'' Spectroscopy}
I would like to show some examples
of the various versions of ``improvement'', 
and remind you of the pictures at the
end of the last chapter to contrast results from  standard actions.

\begin{figure}
\centerline{\ewxy{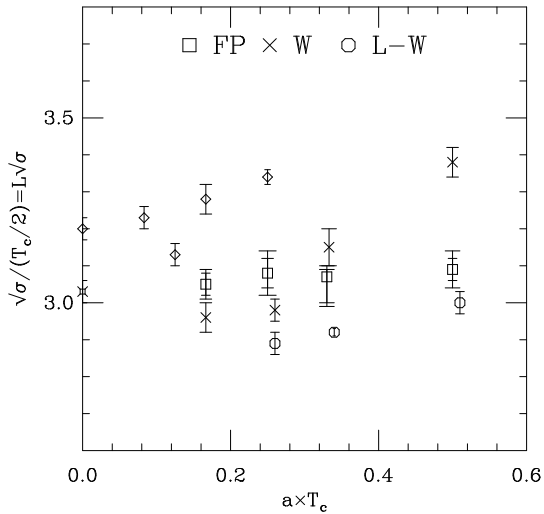}{80mm}
}
\caption{ The (square root of) the string tension in lattices of
constant physical size $L=2/T_c$,
but different lattice spacings (in units of $1/T_c$). 
}
\label{fig:aspect2b}
\end{figure}

Fig. \ref{fig:aspect2b} shows a plot of the string tension measured
in systems of constant physical size (measured in units of $1/T_c$,
the critical temperature for deconfinement), for SU(3) pure gauge theory.
In the quenched approximation, with $\sqrt{\sigma}\simeq 440$ MeV,
$T_c= 275$ MeV and $1/T_c=0.7$ fm.
Simulations with the standard Wilson action are crosses, while the squares show
FP action results  \cite{PAPER123}  and the octagons from the
tadpole-improved L\"uscher-Weisz action  \cite{BLISS}. 
The figure  illustrates that it is hard
to quantify improvement. There are no measurements with the
Wilson action at small lattice spacing of of precisely
the same observables that the ``improvement people'' measured. The best
one can do is to take similar measurements (the diamonds) and attempt
to compute the $a=0$ prediction for the observable we measured (the fancy
cross at $a=0$). This attempt lies on a straight line with the FP
action data, hinting strongly that the FP action is indeed scaling.
The FP action seems to have gained about
a factor of three to four in lattice spacing, or a gain of 
$(3-4)^6$ compared to the 
plaquette action, according to Eq. \ref{COST},
at a cost of a factor of 7 per site because it is more
complicated to program.
The tadpole-improved L\"uscher-Weisz action data lie lower than the
FP action data and do not scale as well.  
As $a\rightarrow 0$ the two actions should yield the same result;
that is just universality at work. However, there is no guarantee that
the approach to the continuum is monotonic.

\begin{figure}
\centerline{\ewxy{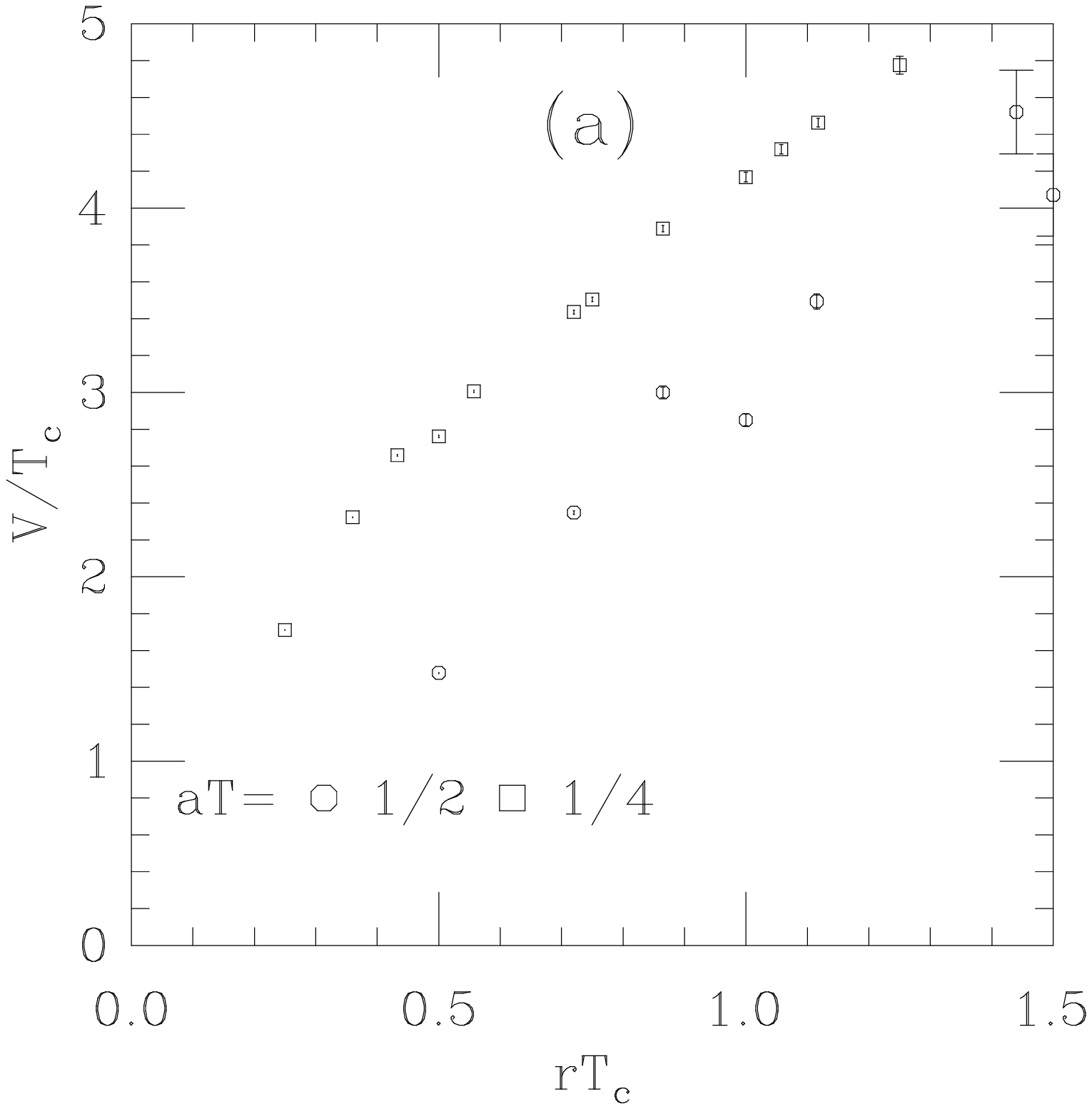}{80mm}
\ewxy{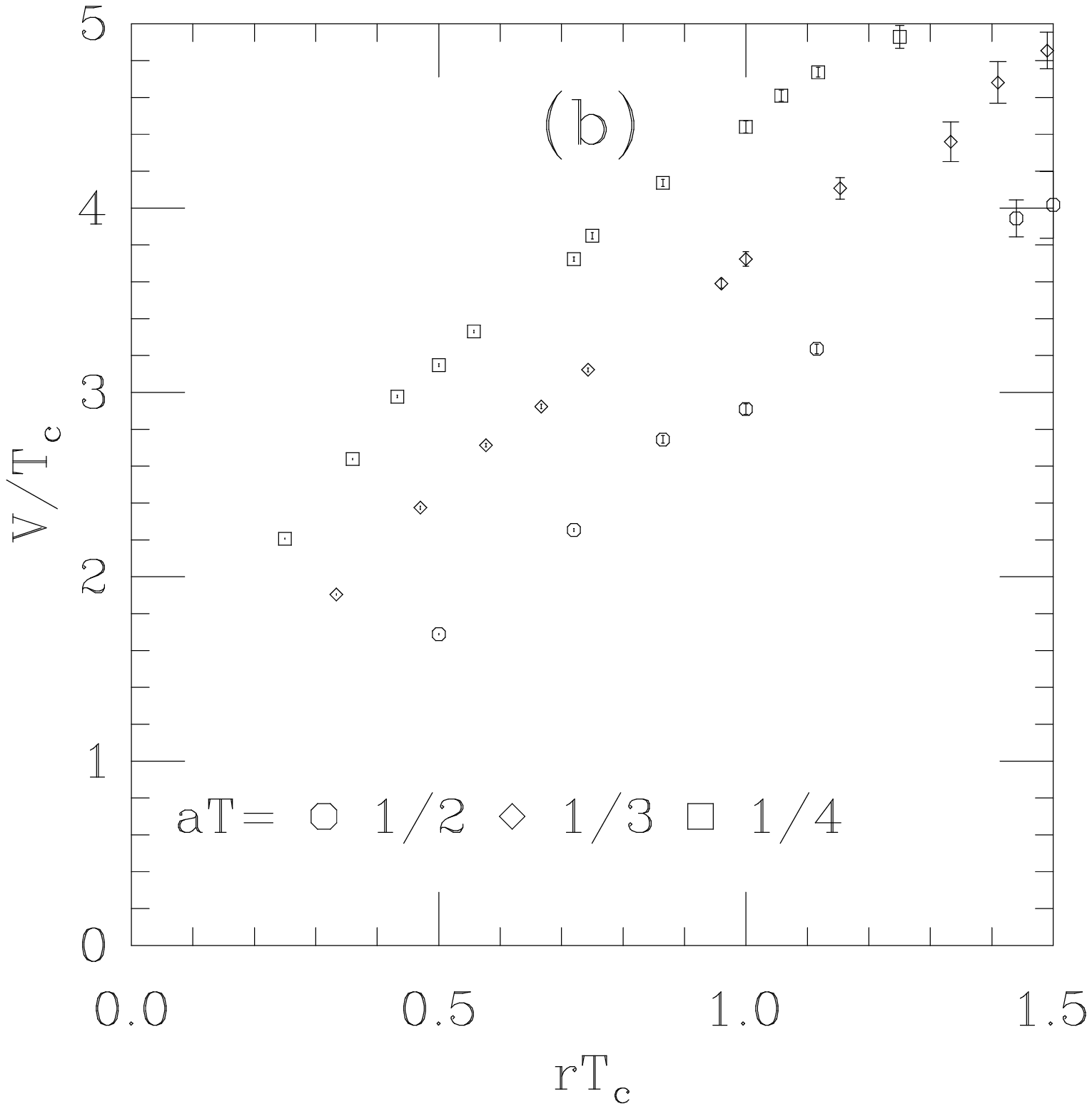}{80mm}}
\caption{ The heavy quark potential in SU(2) pure  gauge theory
measured in units of $T_c$. (a) Wilson action (b) an FP action.}
\label{fig:vr}
\end{figure}

Fig. \ref{fig:vr} shows the heavy quark-antiquark potential in SU(2)
gauge theory, where $V(r)$ and $r$ are measured in the appropriate units
of $T_c$, the critical temperature for deconfinement. The Wilson action
is on the left and a FP action is on the right. The vertical displacements
of the potentials are just there to separate them.
Notice the large violations of rotational symmetry in the Wilson
action data when the lattice spacing is $a= 1/2T_c$ which are
considerably improved in the FP action results.

\begin{figure}
\begin{center}
\setlength{\unitlength}{.02in}
\begin{picture}(120,130)(0,280)
\put(89,340){\circle{3}}\put(95,340){\makebox(0,0)[l]{$0.40$~fm}}
\put(88,329){\framebox(2,2){\mbox{}}}
             \put(95,330){\makebox(0,0)[l]{$0.33$~fm}}
\put(89,320){{\circle*{3}}}\put(95,320){\makebox(0,0)[l]{$0.24$~fm}}
\put(88,310){\rule[-\unitlength]{2\unitlength}{2\unitlength}}
         \put(95,310){\makebox(0,0)[l]{$0.17$~fm}}

\put(15,290){\line(0,1){120}}
\multiput(13,300)(0,50){3}{\line(1,0){4}}
\multiput(14,310)(0,10){9}{\line(1,0){2}}
\put(12,300){\makebox(0,0)[r]{3.0}}
\put(12,350){\makebox(0,0)[r]{3.5}}
\put(12,400){\makebox(0,0)[r]{4.0}}
\put(12,410){\makebox(0,0)[r]{GeV}}

\put(30,290){\makebox(0,0)[t]{$S$}}

\multiput(23,307)(3,0){6}{\line(1,0){2}}
\put(26,307){\circle{3}}
\put(29,306){\framebox(2,2){\mbox{}}}
\put(34,307){\circle*{3}}
\put(37,306){\rule{2\unitlength}{2\unitlength}}

\multiput(23,366)(3,0){6}{\line(1,0){2}}
\put(26,371){\circle{3}}
\put(26,369){\line(0,1){4}}
\put(29,371){\framebox(2,2){\mbox{}}}
\put(30,370){\line(0,1){4}}
\put(34,372){\circle*{3}}
\put(34,368){\line(0,1){8}}
\put(37,369){\rule{2\unitlength}{2\unitlength}}
\put(38,361){\line(0,1){16}}

\put(50,290){\makebox(0,0)[t]{$P$}}

\multiput(43,352)(3,0){6}{\line(1,0){2}}
\put(46,352){\circle{3}}
\put(49,351){\framebox(2,2){}}
\put(54,352){\circle*{3}}
\put(54,351){\line(0,1){2}}
\put(57,351){\rule{2\unitlength}{2\unitlength}}
\put(58,351.5){\line(0,1){1}}

\put(70,290){\makebox(0,0)[t]{$D$}}

\put(66,387){\circle{3}}
\put(66,382){\line(0,1){10}}
\put(69,382){\framebox(2,2){}}
\put(70,376){\line(0,1){12}}
\put(74,387){\circle*{3}}
\put(74,384){\line(0,1){6}}
\put(77,382){\rule{2\unitlength}{2\unitlength}}
\put(78,378){\line(0,1){10}}
\end{picture}


\end{center}
\caption{$S$, $P$, and $D$ states of charmonium computed on lattices with:
$a=0.40$~fm (improved action, $\beta_{plaq}=6.8$);
$a=0.33$~fm (improved action, $\beta_{plaq}=7.1$);
$a=0.24$~fm (improved action, $\beta_{plaq}=7.4$); and
$a=0.17$~fm (Wilson action, $\beta=5.7$, from~[43]),
from Ref. 32. The dashed lines
indicate the true masses.}
\label{fig:spect}
\end{figure}
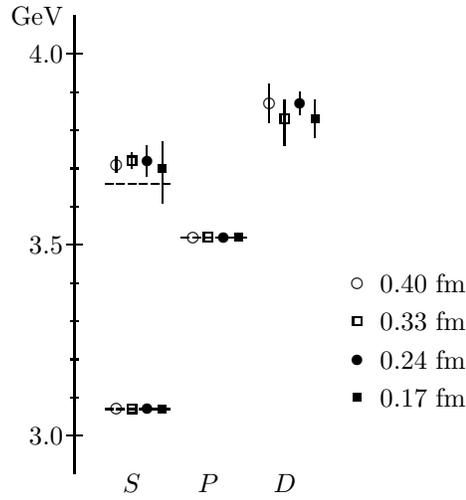

Next we consider nonrelativistic QCD. A comparison of the quenched 
charmonium spectrum from Ref. \cite{PETERIMP} is shown in
Fig. \ref{fig:spect}. When the tadpole-improved L-W action is used to generate
gauge configurations, the scaling window is pushed out to $a\simeq 0.4$
fm for these observables.

\begin{figure}
\centerline{\ewxy{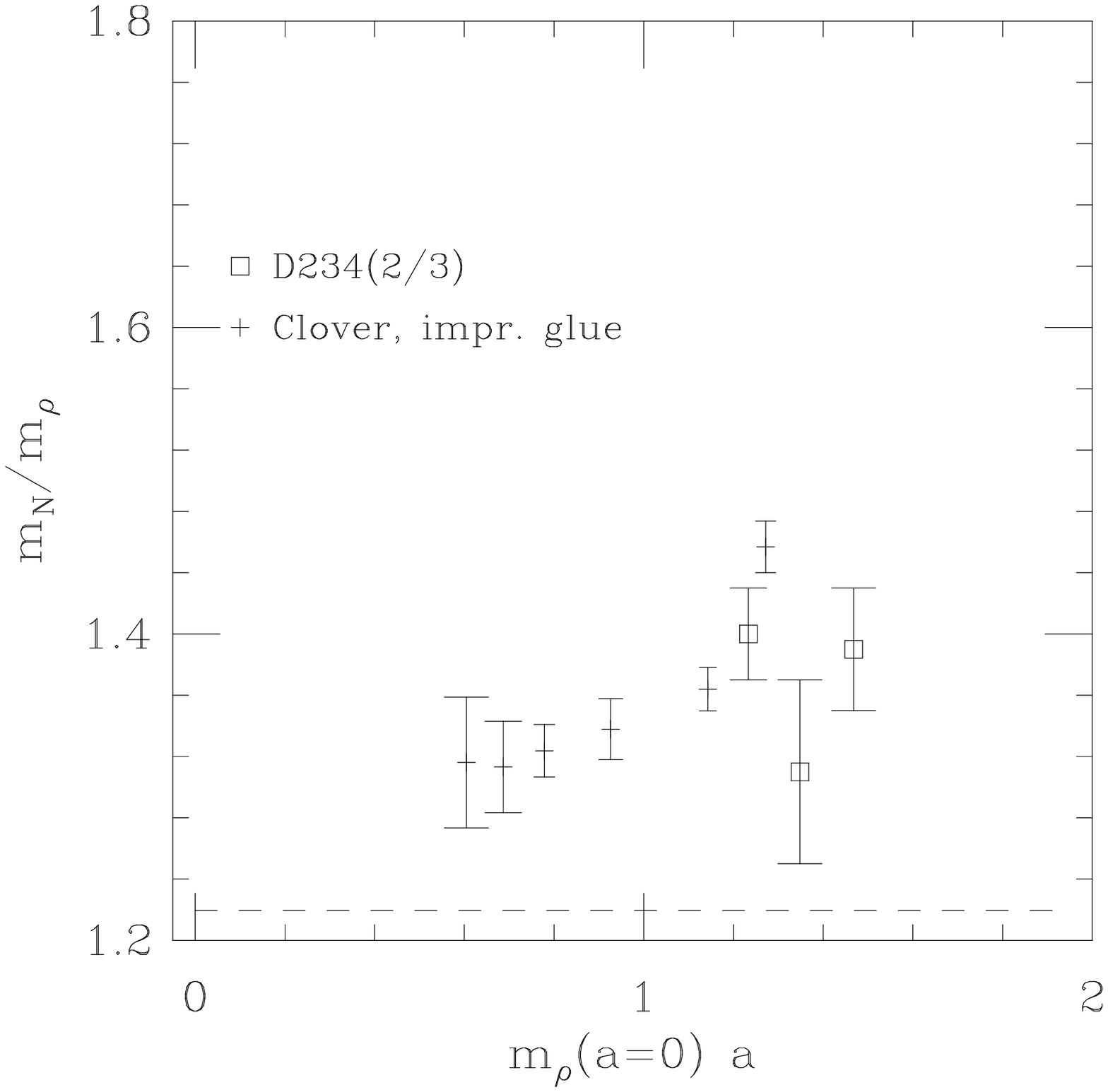}{80mm}
}
\caption{ Nucleon to rho mass ratio (at chiral limit) vs. lattice spacing
(in units of $1/m_\rho$).}
\label{fig:ratiovsmrhoa96}
\end{figure}

\begin{figure}
\centerline{\ewxy{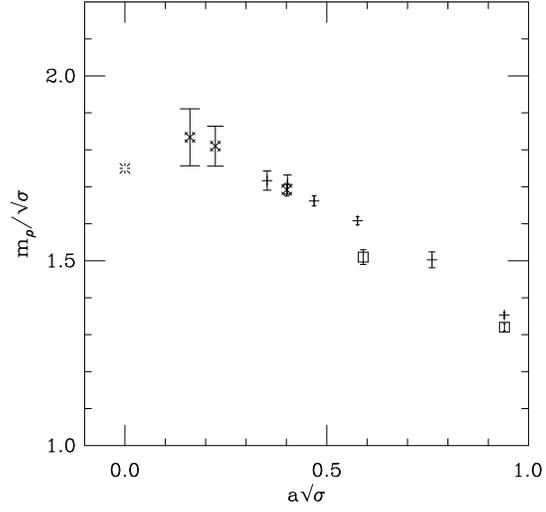}{80mm}
}
%
\caption{Rho mass scaling test with respect to the string tension. }
\label{fig:sommerfigim}
\end{figure}

\begin{figure}
\centerline{\ewxy{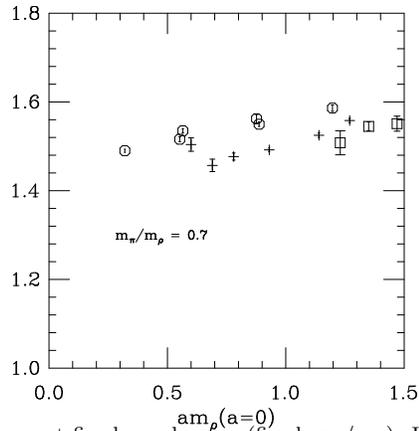}{80mm}
}
\caption{$m_N/ m_\rho$ vs $a m_\rho$ at fixed quark mass
(fixed $m_\pi/m_\rho$).  Interpolations of the S-W and D234(2/3)
data were done by me. }
\label{fig:ratioimp0.7}
\end{figure}

\begin{figure}
\centerline{\ewxy{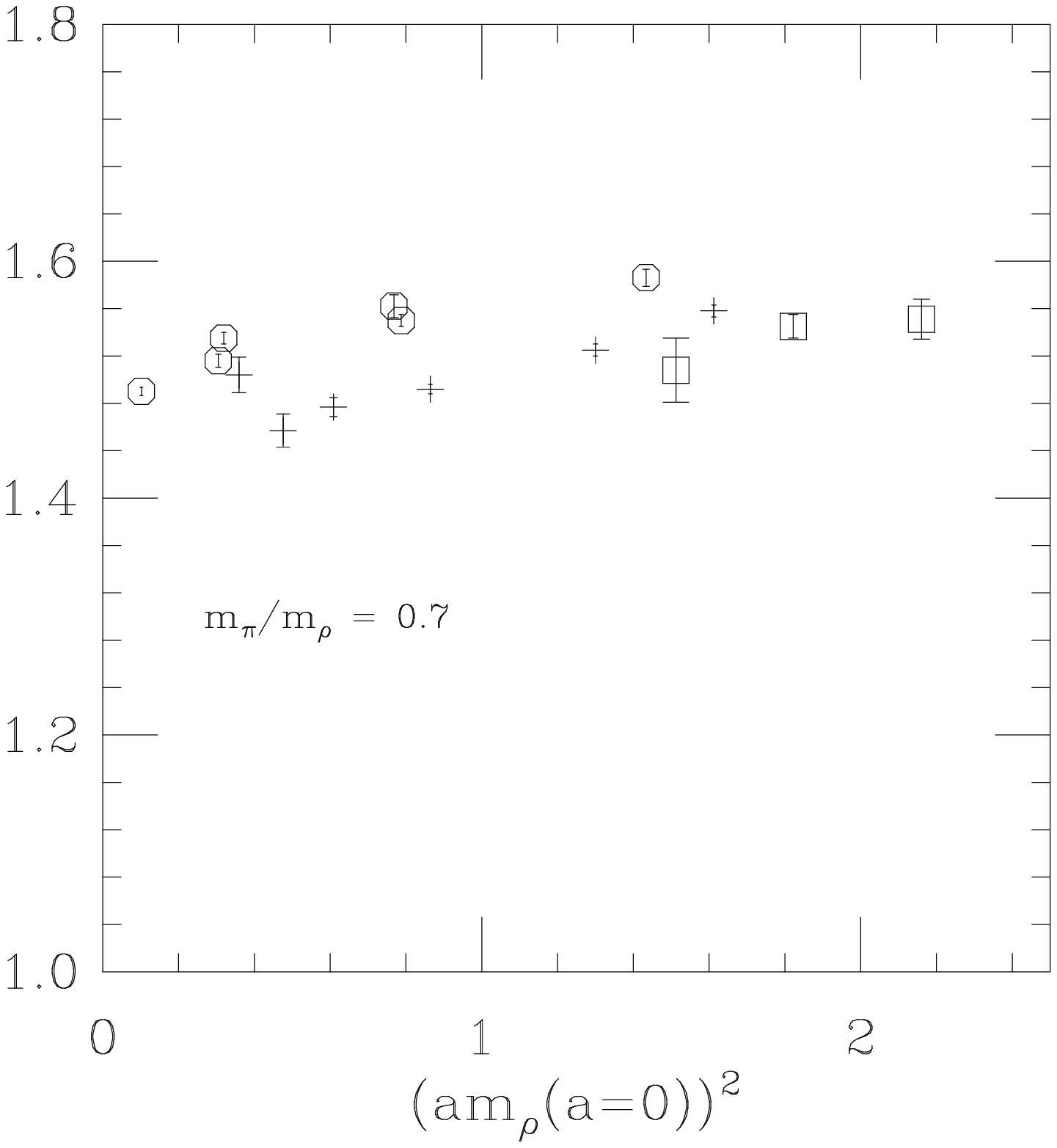}{80mm}
}
\caption{ $m_N/ m_\rho$ vs $(a m_\rho)^2$ at fixed quark mass
(fixed $m_\pi/m_\rho$). }
\label{fig:ratioimp0.7sq}
\end{figure}

Now we turn to tests of quenched QCD for light quarks. 
The two actions which have been most
extensively tested are the S-W action, with and without tadpole improvement,
and an action called the D234(2/3) action, a higher-order variant of
the S-W action  \cite{ALFORD}. Figs. \ref{fig:ratiovsmrhoa96}
and \ref{fig:sommerfigim}, are the analogs of Figs. 
\ref{fig:ratiovsmrhoa} and \ref{fig:sommerfig}.
Diamonds  \cite{UKQCD} and plusses  \cite{SCRI} are S-W actions,
ordinary and tadpole-improved, squares are the D234(2/3) action.
They appear to have about half the scaling violations
as the standard actions but
 they don't remove all scaling violations.
It's a bit hard to quantify the extent of improvement
from these pictures because a chiral extrapolation is hidden in them.
However, one can take one of the ``sections'' of Fig. \ref{fig:fig3combo}
and overlay the new data on it, Fig. \ref{fig:ratioimp0.7}.
It looks like one can double the lattice spacing for an equivalent
amount of scale violation. However, the extrapolation in $a$ is not altogether
clear. Fig. \ref{fig:ratioimp0.7sq} is the same data as Fig.
 \ref{fig:ratioimp0.7}, only plotted vs. $a^2$, not $a$. All of the
actions shown in these figures are supposed to have $O(a^2)$ 
(or better) scaling
violations. Do the data look any straighter in Fig.
\ref{fig:ratioimp0.7sq} than Fig.  \ref{fig:ratioimp0.7}?

\subsection{The bottom line}
At the cost of enormous effort, one can do fairly high
precision simulations of QCD in the quenched approximation with
standard actions. The actions I have shown you appear to
reduce the amount of computation required for
 pure gauge simulations from supercomputers to
very large work stations, probably  a gain of a few hundreds.
All of the light quark data I showed actually came from supercomputers.
According to Eq. \ref{COST}, a factor of 2 in the lattice spacing
gains a factor of 64 in speed. The cost of either
of the two improved actions I showed is about a factor of 8-10
times the fiducial staggered simulation. Improvement methods
for fermions are a few years less mature than ones for pure gauge
theory, and so the next time you hear a talk about the lattice,
things will have changed for the better (maybe).

\section{Case Studies }
\subsection{Glueballs}
To this audience, glueballs are just the closed strings of QCD.
They are interesting because they are part of the QCD spectrum but
not part of a naive quark model.
What do we expect for a spectrum? Models like the bag model, which
have  bound ``constituent gluons,'' would predict that the lightest
state is a scalar, followed by tensor and pseudoscalar states, but
nobody really knows. Here is a place where the lattice is the 
only game in town for a first-principles calculation.

Since this is a TASI for stringy graduate students, the
experimental situation can be summarized briefly: it is a mess
 \cite{TOKI}. There are two leading experimental candidates,
the
 $f_J(1700)$, seen in radiative $\psi$ decays ($\psi \rightarrow
\gamma K\bar K$ and $\psi \rightarrow \gamma \eta\eta$
(mostly at SLAC))
and the $f_0(1500)$, seen in $p\bar p \rightarrow \pi^0\eta\eta$,
from the Crystal Barrel detector at CERN.

People have been trying to measure the masses of the lightest
glueballs (the scalar and the tensor)
using lattice simulations for many years. 
The problem has proven to be very hard, for several reasons.

Recall how we measure a mass from a correlation function (Eqn. \ref{CORRFN}).
The problem with the scalar glueball is that
 the operator $O$ has nonzero vacuum expectation
value, and the correlation function approaches a constant at large $t$:
\bee
\lim_{t \rightarrow \infty}C(t) \rightarrow |\langle 0|O|\vec p = 0\rangle|^2
\exp(-mt) + |\langle 0|O| 0\rangle|^2 .
\ee
The statistical fluctuations on $C(t)$ are given by Eq. \ref{STDEV}
 and we find after a short calculation that
\bee
\sigma \rightarrow {C(0) \over \sqrt{N}} . \ee
Thus the signal to noise ratio collapses at large $t$ like $\sqrt{N} \exp(-mt)$
due to the constant term.
 
A partial cure for this problem is a good trial wave function $O$.
While in principle the plaquette itself could be used, it is so
dominated by ultraviolet fluctuations that it does not produce a good signal.
Instead, people invent ``fat links'' which average the gauge field
over several lattice spacings, and then make interpolating fields
which are closed loops of these fat links. The lattice glueball is
a smoke ring.

\begin{figure}
\centerline{\ewxy{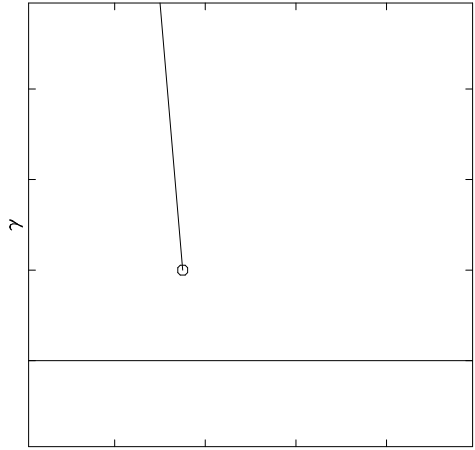}{80mm}
}
\caption{Phase transitions in the fundamental-adjoint plane.}
\label{fig:fundadj}
\end{figure}

The second problem is that lattice actions can have phase transitions
at strong or intermediate coupling, which have nothing to do
with the continuum limit, but mask continuum behavior \cite{BHAN}.
 As an example
of this, consider the gauge group $SU(2)$, where a  link
matrix can be parameterized as
 $U = 1\cos \theta + i \vec \sigma \cdot \vec n \sin \theta$,
so ${\rm Tr}U = 2 \cos \theta$. Now consider a generalization
of the Wilson action $-S = \beta {\rm Tr}U + \gamma({\rm Tr}U)^2$.
(this is a mixed fundamental-adjoint representation action).
At $\gamma \rightarrow \infty$ ${\rm Tr}U \rightarrow \pm 1$ and the
gauge symmetry is broken down to $Z(2)$. But $Z(2)$ gauge theories have
a first order phase transition. First order transitions are stable under
perturbations, and so the phase diagram of this theory, shown
in Fig. \ref{fig:fundadj}, has a line of first order transitions 
which terminate in a second order point. At the second order point some
state with scalar quantum numbers becomes massless. However, now
imagine that you are doing Monte Carlo along the $\gamma=0$ line, that is,
with the Wilson action. When you come near the critical point, any operator
which couples to a scalar particle (like the one you are using to see the 
scalar glueball) will see the nearby transition and the lightest mass
in the scalar channel will shrink. Once you are past the point of
closest approach, the mass will rise again. Any scaling test which ignores
the nearby singularity will lie to you.

\begin{figure}
\centerline{\ewxy{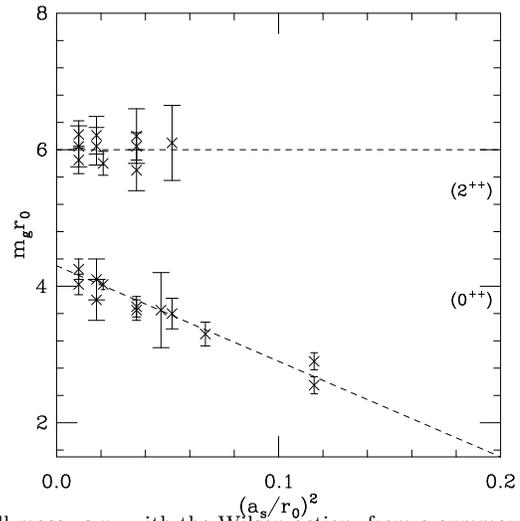}{80mm}
}
\caption{Glueball mass vs $r_0$ with the Wilson action, from a summary
picture in Ref. 49.}
\label{fig:gbr0small}
\end{figure}

This scenario has been mapped out for $SU(3)$, and the place of closest
approach is at a Wilson coupling corresponding to a lattice spacing
of 0.2 fm or so, meaning that very small lattice spacings are needed
before one can extrapolate to zero lattice spacing.
A summary of the situation is shown in Fig. \ref{fig:gbr0small}
 \cite{PEARDON}.
Here the quantity $r_0$ is the ``Sommer radius''  \cite{SOMMER},
defined through the force, by $r_0^2F(r_0)= -1.65$.
In the physical world of three colors and four flavors, $r_0 = 0.5$ fm.

Finally, other arguments suggest that a small lattice spacing or
a good approximation to an  action on an RT are needed to for
glueballs:  the physical diameter of the glueball, as inferred from
the size of the best interpolating field, is small, about 0.5 fm.
 Sh\"afer and Shuryak  \cite{SS} have argued that the small size is due to
instanton effects.
Most lattice actions at large lattice spacing
do bad things to instantons  \cite{INSTANTON12}.

Two big simulations have carried calculations of the glueball mass
close to the continuum limit: the UKQCD collaboration  \cite{UKQCDGB}
and a collaboration at IBM which built its own computer  \cite{GF11}.
(The latter group is the one with the press release last December
announcing the discovery of the glueball.)
Their predictions in MeV are different and they each favor a different
experimental candidate for the scalar glueball (the one which is
closer to their prediction, of course). It is a useful object lesson
because both groups agree that their lattice numbers agree before
extrapolation, but they extrapolate differently to $a=0$.

The UKQCD group sees that the ratio $m(0^{++})/\surd\sigma$ can be
well fitted with  a form   $b + c a^2 \sigma$
($\sigma$ is the string tension) and a fit of this form to the lattice
data of both groups gives
$ m(0^{++})/\surd\sigma = 3.64 \pm 0.15$.
To turn this into MeV we need $\sigma$ in MeV units. One way
is to take $m_{\rho}/\surd\sigma$ and extrapolate that
to $a=0$ using $ b + c a \surd\sigma$. 
Averaging and putting 770 MeV for $m_{\rho}$ one gets
$\surd\sigma = 432 \pm 15$ MeV, which is  consistent
with the usual estimate (from extracting the string tension
from the heavy quark potential)
of about 440 MeV.
Using the total average they get 
$m(0^{++}) = 1572 \pm 65 \pm 55$ MeV
where the first error is statistical and the second comes
from the scale. 

The IBM group, on the other hand, notices that $m_\rho a$ and $m_\phi a$
scale asymptotically, use the phi mass to predict 
quenched $\Lambda_{\bar{MS}}$,
then extrapolate $m(0^{++})/\Lambda = A + B(a \Lambda)^2$. They get
1740(41) MeV from their data,  when they analyze UKQCD data, they
get 1625(94) MeV, and when they combine the data sets, they get 1707(64)
MeV.

A neutral reporter could get hurt here. It seems to me that the lattice  prediction for the scalar glueball
is $1600 \pm 100 $ MeV, and that there are two experimental candidates
for it. 

Masses are not the end of the story. The IBM group has done two 
interesting recent calculations related to glueballs, which strengthen their
claim that the $f_J(1710)$ is the glueball.

\begin{figure}
\centerline{\ewxy{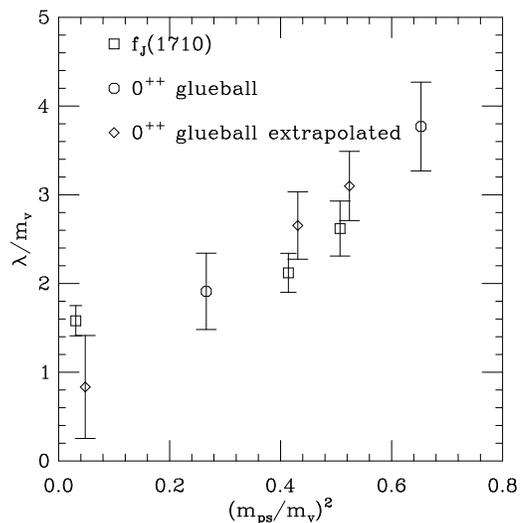}{80mm}
}
\caption{Scalar glueball decay couplings from Ref. 53.}
\label{fig:gbdecay}
\end{figure}

The first one  of them  \cite{GF11DE} was actually responsible for the
press release. It is a calculation of the decay width of the
glueball into pairs of pseudoscalars. This is done by computing
an unamputated three point function on the lattice, with an
assumed form for the vertex, whose magnitude is fitted. The result
 is shown in Fig. \ref{fig:gbdecay}. 
The octagons  are the results of the simulation and
the diamonds show interpolations in the quark mass.
The ``experimental'' points (squares) are from
a partial wave analysis of isoscalar scalar resonances by Longacre
and Lindenbaum  \cite{LL}.

The response of a member of the other side is that
the slope of the straight line that one would put through
the three experimental points is barely, if at all, compatible with
the slope of the theoretical points. Since they argue
theoretically for a straight line, the comparison
of such slopes is a valid one. 

If one of the experimental states is not a glueball, it is likely
to be a ${}^3P_0$ orbital excitation of quarks.
 Weingarten and Lee  \cite{GF11P0} are computing the mass
of this state on the lattice and argue that it is lighter than
1700 MeV; in their picture the $f_0(1500)$ is an $s \bar s$ state. 
I have now said more than I know and will just refer you to
  recent discussions of the question  \cite{GBFIGHT}.

Both groups predict that the $2^{++}$ glueball is at about 2300 MeV.

\begin{figure}
\centerline{\ewxy{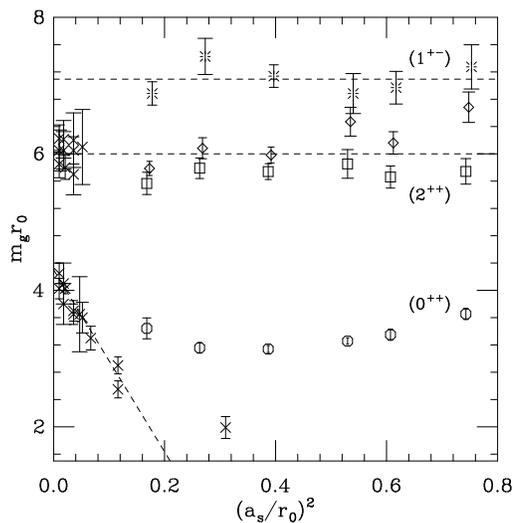}{80mm}
}
\caption{Glueball mass vs $r_0$ from Ref. 49, including large
lattice spacing data.}
\label{fig:gbr0}
\end{figure}

Can ``improved actions'' help the situation? Recently, Peardon and Morningstar
 \cite{PEARDON}
implemented a clever method for beating the exponential signal-to-noise ratio:
make the lattice spacing smaller in the time direction than in the space direction. Then the signal, which falls like $\exp (-m a_t L_t)$ after
$L_t$ lattice spacings, dies more slowly because $a_t$ is reduced.
Their picture of the glueball mass vs $r_0$. is shown in Fig. \ref{fig:gbr0}.
They are using the tadpole-improved L\"uscher-Weisz action.
The pessimist notes the prominent dip in the middle of the curve; this
action also has a lattice-artifact transition (somewhere); the optimist
notes that the dip is much smaller than for the Wilson action and
then the pessimist notes that there is no Wilson action data at large
lattice spacing to compare. I think the jury is still out.

\subsection{$\alpha_s(M_Z)$}

For some time now there have been claims that physics at the Z pole
hints at a possible breakdown in the standard model  \cite{SHIFMAN}.
A key question in the discussion is whether or not
 the value of $\alpha_{\bar {MS}}$ inferred from the decay width of the
Z is anomalously high relative to other determinations of the strong
coupling (when run to the Z pole, usually).

The most recent analysis I am aware of is due to
Erler and Langacker  \cite{PDGALPHA}.
Currently, $\alpha_{\overline{MS}}^{lineshape}=  0.123(4)(2)(1)$ for the standard
model Higgs mass range, where the first/ second/ third uncertainty
is from inputs/ Higgs mass/ 
estimate of $\alpha_s^4$ terms. The central Higgs mass
is assumed to be 300 GeV, and the second error is for $M_H=1000$  GeV (+),
 60 GeV (-).
For the SUSY Higgs mass range (60-150 GeV), one has the lower value 
$\alpha_{\overline{MS}}=.121(4)(+1-0)(1)$. A global fit to all data gives
0.121(4)(1). Hinchcliffe in the same compilation quotes a global average
of 0.118(3).

The lattice can contribute to this question by predicting 
$\alpha_{\overline{MS}}$ from low energy physics. The basic idea is simple:
The lattice is a (peculiar) UV cutoff. A lattice mass $\mu = Ma$ plus
an experimental mass $M$ give a lattice spacing $a = \mu/M$ in fm.
If one can measure some quantity related to $\alpha_s$ at a scale
$Q\simeq 1/a$, one can then run the coupling constant out to the Z.

\begin{figure}
\centerline{\ewxy{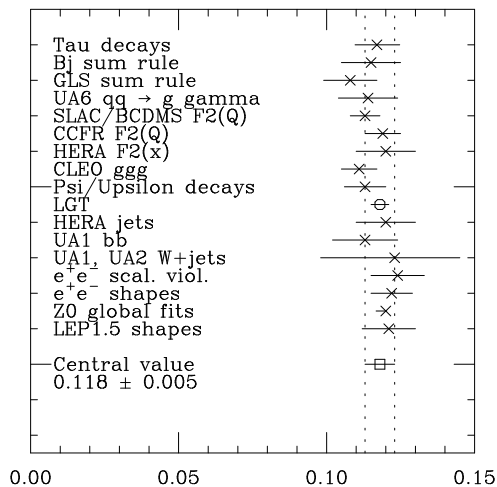}{80mm}
}
\caption{ Survey of $\alpha_{\overline{MS}}(M_Z)$ from Ref. 61.}
\label{fig:allalpha}
\end{figure}

The best (recent) lattice number, from Shigemitsu's Lattice 96
summary talk  \cite{SHIGEMITSU}, is
\bee
\alpha_{\overline{MS}}(Z) = 0.1159(19)(13)(19)
\ee
where the first error includes both statistics and estimates of
discretization errors, the second is due to uncertainties from the
dynamical quark mass, and the third is from conversions of conventions.
The lattice number is about one standard deviation below the pure
Z-physics number. Lattice results are compared
to other recent determinations of $\alpha_{\overline{MS}}(Z)$
 in Fig. \ref{fig:allalpha}, a figure provided by P. Burrows \cite{BURROWS}.

Two ways of calculating $\alpha_s(M_Z)$ from lattice have been proposed:
The first is the ``small loop method'' 
  \cite{AIDA}. This method uses the ``improved
perturbation theory'' described in  Chapter 3:
One assumes that a version of perturbation theory can describe the behavior
of short distance objects on the lattice: in particular, that the
plaquette can be used to define $\alpha_V(q=3.41/a)$. With typical
lattice spacings now in use, this gives the coupling at a momentum 
$Q_0=8-10$ GeV.  One then converts the coupling to $\alpha_{\overline{MS}}$
and runs out to the Z using the (published) three-loop beta function
 \cite{ROD}.

Usually, the lattice spacing is determined from the mass splittings
of heavy $Q \bar Q$ states. This is done because the mass differences between
physical heavy quark states are nearly independent of the quark mass--
for example, the S-P mass splitting of the $\psi$ family is about 460
MeV, and it is about 440 MeV for the $\Upsilon$. A second reason is that
the mass splitting is believed to be much less sensitive to sea
quark effects than light quark observables, and one can estimate
the effects of sea quarks through simple potential models.
The uncertainty in the lattice spacing is three to five per cent,
but systematic effects are much greater (as we will see below).

The coupling constant comes from Eq. \ref{PLAQALPHA}. The plaquette can be
measured to exquisite accuracy (0.01 per cent is not atypical) and so the 
coupling constant is known essentially without error. However, the
scale of the coupling is uncertain (due to the lattice spacing).

The next problem is getting from lattice simulations, which are done with
$n_f=0$ (quenched) or $n_f=2$ (but unphysical sea quark masses)
to the real world of $n_f=3$. Before simulations with dynamical fermions
were available, the translation was done by running down in $Q$ to
a ``typical gluonic scale'' for the psi or the upsilon (a few hundred MeV)
and then matching the coupling to the three-flavor coupling (in the spirit
of effective field theories). Now we have simulations at $n_f=2$.
Recall that in lowest order
\bee
{1 \over \alpha_s} = \big( {{11 - {2 \over 3}n_f} \over {4\pi}}  \big) 
\ln {Q^2 \over \Lambda^2}
\label{ALINV}
\ee
One measures $1/\alpha_s$ in two simulations, one quenched, the other
at $n_f=2$, runs one measurement in $Q$ to the $Q$ of the other,
then extrapolates $1/\alpha$ linearly in $n_f$ to $n_f=3$.
Then one can convert to $\bar{MS}$ and run away.

\begin{figure}
\centerline{\ewxy{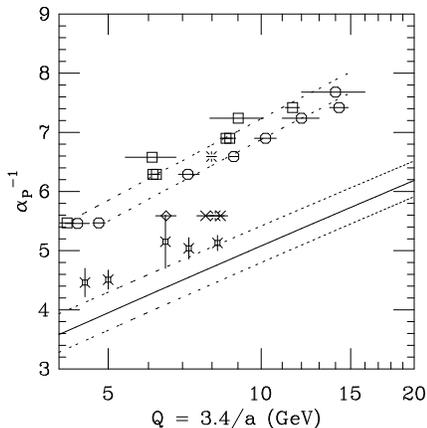}{80mm}
}
\caption{ Survey of $\alpha_{\overline{MS}}(Q)$ at the scale where lattice
simulations are actually done.}
\label{fig:alphap2js}
\end{figure}

Pictures like Fig. \ref{fig:allalpha} are not very useful
when one wants to get a feel for the errors inherent in the lattice
calculation. Instead, let's run our expectations for $\alpha_s(M_Z)$
down to the scale where the lattice simulations are done, and compare.
Fig. \ref{fig:alphap2js} does that. The squares are the results of
simulations of charmed quarks and the octagons are from bottom
quarks, both with $n_f=0$. The crosses and diamond are $n_f=2$
bottom and charm results. 
(The bursts show upsilon data when the 1S-2S mass difference gives a 
lattice spacing.) 
Note the horizontal error bars on the lattice data.
Finally, the predicted $n_f=3$ coupling $\alpha_P$ is shown
in the fancy squares, with error bars now rotated because the
convention is to quote an error in $\alpha_s$.  The lower three lines
in the picture (from top to bottom) 
are $\alpha_{\overline{MS}}(M_Z)=0.118$, 0.123, and 0.128
 run down and converted to the lattice prescription.

The two top lines are predictions for how quenched $\alpha$ should run.

Now for the bad news. All of the $n_f=2$ data shown here were actually 
run on the same set of configurations. The bare
couplings are the same, but the
lattice spacings came out different. 
What is happening is that we are taking  calculations at some lattice
spacing and inferring a continuum numbers from them, but the lattice
predictions have scale violations which are different.
(The $\Upsilon$ calculations use nonrelativistic quarks, the
$\psi$ calculations use heavy Wilson quarks.)
Notice also that the bottom and
charm quenched lattice spacings are different. This discrepancy
is thought to be a failure of the quenched approximation: the characteristic
momentum scale for binding in the $\psi$ and $\Upsilon$ are different,
and because $n_f$ is not the real world value $\alpha$ runs incorrectly
between the two scales.
Said differently, in the quenched approximation, the spectrum of
heavy quark bound states is different from the real world.

There is a second method of determining a running coupling constant which
actually allows one to see the running over a large range of scales.
It goes by the name of the ``Schr\"odinger functional,''
 (referring to the
fact that the authors study QCD in a little box with specified
boundary conditions) but
``coupling determined by varying the box size'' would be a more descriptive
title. It has been applied to quenched QCD but has not yet been
extended to full QCD. I will describe the method in the
context of the $d=2$ sigma
model  \cite{SCHSIG} rather than QCD  \cite{SCHQCD}.

The idea is that coupling constants can be defined through the response
of a system to boundary conditions. For example, if
one dimension of a $d=2$ $O(n)$ sigma model is compactified with
size $L$, the
mass gap $M(L)$
(defined through the transfer matrix along the other dimension)
is related to the coupling through
\bee
g^2(L)= 2M(L)L/(n-1).
\ee
The coefficients insure that the answer reduces to the lowest-order
result.
Now we take take the beta function
\bee
\beta(g)  =  L {{dg^2}\over{dL}},
\ee
stretch the length of the compact dimension from $L_0$ to $sL_0$,
and compute the change in the coupling
\bee
\int_{L_0}^{sL_0} {{dL}\over L} =  \int_{g^2(L_0)}^{g^2(sL_0)}
{{dg^2}\over{\beta(g^2)}} \equiv \int_u^{\sigma(s,u)}{{dv}\over{\beta(v)}}
\ee
where the ``step scaling function'' $\sigma(s,u=g^2(L))=g^2(sL)$
is the new coupling constant.

Now the idea is to do simulations with the same bare coupling on
systems of size $L_0$ and $sL_0$ and, by measuring $u$ and $\sigma(s,u)$,
to see how the new coupling depends on the original one.
An example of how this works is shown in Table \ref{tab:tabsch1}.
Each horizontal pair are the coupling and its step function 
(for $s=2$). The output coupling on the first line is used as the input
coupling on the second line (or rather, since these couplings are derived,
not bare couplings, one must interpolate to begin with the old output
coupling as the new input coupling). When the matching is not perfect,
perturbation theory is used for the small amount of running which is required.
After four steps the physical length scale has 
increased by a factor of $2^4=16$.

\begin{table}[t]\caption{Data for the running coupling constant in pure SU(2) gauge theory. \label{tab:tabsch1}}
\vspace{0.4cm}
\begin{center}
\begin{tabular}{|c|l|}
\hline &  \\ u & $\sigma(2,u)$ \\
\hline
2.037 & 2.45(4) \\
2.380 & 2.84(6) \\
2.840 & 3.54(8) \\
3.55 & 4.76 \\
\hline
\end{tabular}
\end{center}
\end{table}

Now we can unfold the couplings. We measure distances in terms of the
largest $L$ in the simulation and show 
the coupling constant in Table \ref{tab:tabsch2}.

\begin{table}[t]\caption{The Sc\"hrodinger functional
running coupling constant in pure SU(2) gauge theory. \label{tab:tabsch2}}
\vspace{0.4cm}
\begin{center}
\begin{tabular}{|c|l|}
\hline &  \\ $L/L_{max}$ & $g^2(L)$ \\
\hline
1.0 & 4.765 \\
0.500(23) & 3.55 \\
0.249(19) & 2.840 \\
0.124(13) & 2.380 \\
0.070(8) & 2.037 \\
\hline
\end{tabular}
\end{center}
\end{table}

\begin{figure}
\centerline{\ewxy{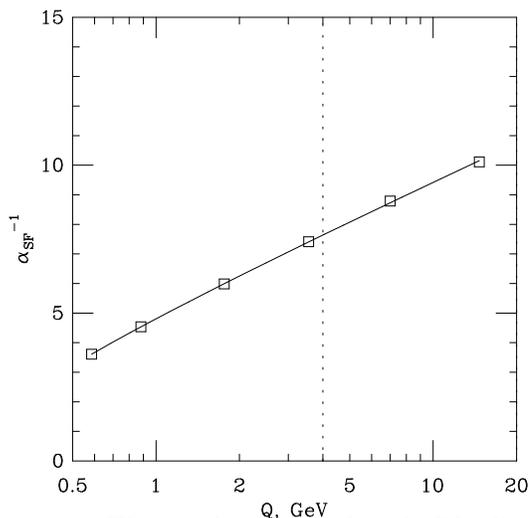}{80mm}
}
\caption{ The pure gauge SU(3) coupling constant from the
Schrodinger functional method of Ref. 65, with superimposed three-loop
prediction. The data from Fig. 29 span the range to the right of
the dotted line.}
\label{fig:alphasf}
\end{figure}

We can introduce a scale in GeV by measuring something  physical
with the bare
parameters corresponding to one of the $L$'s, say, for the gauge
theory, the Sommer
variable  $r_0$ (it should be obvious why the authors use this observable)
at $L_{max}$. An example of a running coupling constant via
this prescription is shown in Fig.  \ref{fig:alphasf}.
This is a coupling constant in a particular prescription; when it
is small enough it could be matched to any other prescription using
perturbation theory.

All the simulations are done in small lattice volumes.
I have left out many technical details, of course.
Please note that this calculation checks perturbation theory; it
does not use it overtly (apart from the small amount of running I
described). I think it is beautiful  \cite{WEISZAL}.

\section{Conclusions}
There it is! I now appreciate what the ``string lecturer'' goes through
when he arrives at TASI's focussed on phenomenology.
I have left out a lot. For example, the lattice is a fertile source
of numbers for comparisons with the standard model, where what the 
experimentalist measures is often a combination of a ``fundamental
parameter'' of the standard model times a hadronic matrix element.
The lattice and lattice QCD in particular are going through a lot of changes
right now, and perhaps for the better.
I  hope that some of the methods I have described might be
useful as a starting point when you find yourselves working in the
strong-coupling sector of your favorite theory.

\section*{Acknowledgements}
I would like to thank 
M.~Alford,
P.~Burrows,
S.~Gottlieb,
A.~Hasenfratz,
P.~Hasenfratz,
U.~Heller,
P.~Langacker,
P.~Lepage,
P.~Mackenzie,
J.~Negele,
F.~Niedermayer,
J.~Shigemitsu,
J.~Simone,
R.~Sommer,
R.~Sugar,
M.~Teper,
D.~Toussaint,
D.~Weingarten,
U.~Wiese,
and
M.~Wingate
for discussions, figures, and correspondence.
They have all influenced these lectures, but probably not in the direction
that they intended.
I would also like to thank the Institute for Theoretical Physics
at the University of Bern for its hospitality, where these lectures
were written.
This work was supported by the U.~S. Department of Energy.

\newcommand{\PL}[3]{{Phys. Lett.} {\bf #1} {(19#2)} #3}
\newcommand{\PR}[3]{{Phys. Rev.} {\bf #1} {(19#2)}  #3}
\newcommand{\NP}[3]{{Nucl. Phys.} {\bf #1} {(19#2)} #3}
\newcommand{\PRL}[3]{{Phys. Rev. Lett.} {\bf #1} {(19#2)} #3}
\newcommand{\PREPC}[3]{{Phys. Rep.} {\bf #1} {(19#2)}  #3}
\newcommand{\ZPHYS}[3]{{Z. Phys.} {\bf #1} {(19#2)} #3}
\newcommand{\ANN}[3]{{Ann. Phys. (N.Y.)} {\bf #1} {(19#2)} #3}
\newcommand{\HELV}[3]{{Helv. Phys. Acta} {\bf #1} {(19#2)} #3}
\newcommand{\NC}[3]{{Nuovo Cim.} {\bf #1} {(19#2)} #3}
\newcommand{\CMP}[3]{{Comm. Math. Phys.} {\bf #1} {(19#2)} #3}
\newcommand{\REVMP}[3]{{Rev. Mod. Phys.} {\bf #1} {(19#2)} #3}
\newcommand{\ADD}[3]{{\hspace{.1truecm}}{\bf #1} {(19#2)} #3}
\newcommand{\PA}[3] {{Physica} {\bf #1} {(19#2)} #3}
\newcommand{\JE}[3] {{JETP} {\bf #1} {(19#2)} #3}
\newcommand{\FS}[3] {{Nucl. Phys.} {\bf #1}{[FS#2]} {(19#2)} #3}


\begin{thebibliography}{99}
\bibitem{DOUGSHOW}
This particular formulation of the problem was shown to me by D. Toussaint.
\end{thebibliography}
\end{document}